\documentclass[useAMS,usenatbib,usegraphicx]{mn2e}
\usepackage{psfig}   
\usepackage{graphicx}
\usepackage{amssymb}
\usepackage{subfigure}
\usepackage{longtable}

\title[Supernova enrichment of G-dwarfs]{Supernova enrichment and dynamical histories of solar-type stars in clusters}

\author[R. J. Parker, R. P. Church, M. B. Davies \& M. R. Meyer]{Richard J.~Parker$^1$\thanks{E-mail: rparker@phys.ethz.ch}, Ross P. Church$^2$, Melvyn B. Davies$^2$ and Michael R. Meyer$^1$ \vspace*{0.1cm}\\
   $^1$Institute for Astronomy, ETH Z{\"u}rich, Wolfgang-Pauli-Strasse 27, 8093, Z{\"u}rich, Switzerland\\
   $^2$Lund Observatory, Department of Astronomy and Theoretical Physics, Box 43, SE-221 00 Lund, Sweden}

\begin{document}

\date{}
                             
\pagerange{\pageref{firstpage}--\pageref{lastpage}} \pubyear{2013}

\maketitle

\label{firstpage}

\begin{abstract}
We use $N$-body simulations of star cluster evolution to explore the hypothesis that short-lived radioactive isotopes found in meteorites, such as $^{26}$Al, were delivered to the 
Sun's protoplanetary disc from a supernova at the epoch of Solar System formation. We cover a range of star cluster formation parameter space and model  
both clusters with primordial substructure, and those with smooth profiles. We also adopt different initial virial ratios -- from cool, collapsing clusters 
to warm, expanding associations. In each cluster we place the same stellar population; the clusters each have 2100 stars, and contain one massive 
25\,M$_\odot$ star which is expected to explode as a supernova at about 6.6\,Myr. We determine the number of Solar (G)-type stars that are within 0.1 -- 0.3\,pc of the  25\,M$_\odot$ star at the time of the supernova, which is the distance required to enrich 
the protoplanetary disc with the $^{26}$Al abundances found in meteorites. We then determine how many of these G-dwarfs are unperturbed `singletons'; stars which 
are never in close binaries, nor suffer sub-100\,au encounters, and which also do not suffer strong dynamical perturbations.

The evolution of a suite of twenty initially identical clusters is highly stochastic, with the supernova enriching over 10 G-dwarfs in some clusters, and none at all in others. 
Typically only $\sim$25\,per cent of clusters contain enriched, unperturbed singletons, and usually only 1 -- 2 per cluster (from a total of 96 G-dwarfs in each cluster). 
The initial conditions for star formation do not strongly affect the results, although a higher fraction of supervirial (expanding) clusters would contain enriched G-dwarfs
if the supernova occurred earlier than 6.6\,Myr. If we sum together simulations with identical initial conditions, then $\sim$1\,per cent of all G-dwarfs in our simulations are enriched, unperturbed singletons.
\end{abstract}

\begin{keywords}   
stars: formation -- kinematics and dynamics -- open clusters and associations: general -- planetary systems -- methods: numerical

\end{keywords}

\section{Introduction}

One of the outstanding issues in astrophysics is characterising the birth environment of the Solar System \citep[e.g.][]{Adams11}. In particular, 
understanding whether the Sun is an `average' star in terms of its formation and evolution is important for assessing how likely the formation 
of a quiescent, habitable Solar System is when placed in the context of other planetary systems. 

A strong constraint on the formation of our Solar System appears to be the presence of short-lived radioactive isotopes in meteorites originating 
from the epoch of planet formation \citep{Lee76}. The short half-life and abundance of such isotopes (inferred from their stable daughter products) argues for their rapid inclusion in meteorites 
during the early phases of the Solar System \citep{Looney06,Thrane06}.

Short-lived radiogenic isotopes may also be the dominant heat source for forming planetesimals in protoplanetary discs \citep{Urey55,MacPherson95}. This could affect the 
survival of volatile elements in the inner region of the Solar System and have implications for planet habitability \citep{Nimmo02}.

Several short-lived isotopes with half-lives ranging from tens of days to several Myr are present in meteorites, but two -- $^{26}$Al and $^{60}$Fe -- are very difficult to produce 
without nucleosynthesis in massive stars \citep[e.g.][]{Goswami04}. It is possible to produce $^{26}$Al through spallation \citep{Lee98,Shu01} or from evolved asymtotic giant branch (AGB) stars \citep{Busso99,Busso03}, but the presence of $^{60}$Fe 
points toward enrichment from a supernova explosion \citep[see e.g.\,\,the discussion in][]{Adams11}\footnote{Note that recent work \citep[e.g.][]{Moynier11,Tang12} has suggested that the abundance of $^{60}$Fe in the early Solar System may not be as high as previously thought, 
and could be as low as the levels measured in the background interstellar medium.}.

Following the discovery of $^{26}$Al in meteorites, \citet{Cameron77} suggested that the Sun could have formed when a supernova explosion triggered the collapse of a 
star-forming giant molecular cloud \citep[GMC, see also][]{Boss95,Cameron95,Boss00,Boss12}. This scenario requires that the supernova explosion does not destroy the GMC 
altogether.

Other authors have suggested that whereas $^{60}$Fe may be delivered from a supernova, $^{26}$Al can also be produced in the winds of evolved massive stars \citep{Gounelle12}, 
and that the isotope enrichment occurs in a sequential star formation process. Firstly, $^{60}$Fe is delivered to the nearby interstellar medium (ISM) by multiple supernovae from 
the first generation of star formation. These supernovae then trigger a second generation of star formation in which $^{26}$Al is delivered into the ISM by the wind of a single 
massive star. The Sun is then born in a third generation of star formation within the shell of contaminated ISM material.

Finally, $^{26}$Al and $^{60}$Fe can be delivered directly to the disc from which the Solar System formed \citep{Chevalier00,Ouellette07}. In this scenario, the massive star is 
assumed to form coevally with the Sun, but it evolves faster and the resultant core collapse supernova occurs before the protoplanetary disc has begun to coalesce and form large planetesimals.  
To obtain the correct enrichment levels, \citet{Chevalier00} and \citet{Ouellette07} suggest that the Sun's protoplanetary disc must have been between 0.1 and 0.3\,pc from the 
supernova (at distances less than 0.1\,pc, the supernova is likely to strip away too much of the disc, and beyond 0.3\,pc the yield of radioactive isotopes is too low, \citealp{Adams11}).

If the meteorite enrichment occurs during a single supernova explosion, then a 25\,M$_\odot$ star is most likely to deliver the relative isotopic abundances \citep{Wasserburg06}. 
At first sight, a 25\,M$_\odot$ star in close proximity to the Sun may seem unlikely; most stars form in clusters or associations \citep{Lada03} and there is a relation between 
the most massive star that can form in a cluster and the mass of the cluster (cf. number of stars) itself \citep{Weidner13}. \citet{Adams01} show that when randomly sampling 
an initial mass function (IMF) a 25\,M$_\odot$ star is likely to form in a cluster with at least $N = 2000 \pm 100$ other stars (the exact number of stars depends on the adopted IMF). 

This moderately high expectation value for the number of stars that form in the company of a 25\,M$_\odot$ star, coupled with the fact that embedded clusters typically have radii 
less than several pc \citep{Lada03}, suggests that the birth environment of the Solar System could be rather dense and therefore hostile. UV radiation from massive stars, which would evaporate 
or truncate the protoplanetary disc \citep{Armitage00,Scally01,Adams04}, and dynamical interactions during close encounters with intermediate and low-mass stars \citep{Bonnell01b,Adams06,Parker12a} 
could inhibit planet formation in such an environment. 

Several authors have estimated the maximum number of stars in the Sun's natal cluster that would allow the formation of a 25\,M$_\odot$ star, but also not be too hostile for the 
formation and evolution of the Solar System. For example, \citet{Adams01,Adams06,Dukes12} calculate collisional cross sections for the Solar System to undergo disruptive 
interactions with passing stars and \citet{Pfalzner13} calculates the likely encounter rates for Sun-like stars in two different types of star forming region; an extremely dense 
cluster versus a more diffuse OB association. In general these authors find that a cluster with $N = 10^3 - 10^4$ stars would enable the formation of a 25\,M$_\odot$ star without 
dynamical interactions prohibiting the formation of the Solar System -- provided that the cluster quickly disperses \citep{Dukes12,Pfalzner13}.

However, it remains unclear whether the evolution of a `typical' cluster which forms a  25\,M$_\odot$ star does result in supernova enrichment of G-dwarf stars like the Sun, without 
those G-dwarfs suffering dynamical interactions which would hinder or disrupt planet formation. Previous work on this topic has assumed that if the cluster contains a 25\,M$_\odot$ star then 
enrichment is virtually guaranteed, and instead focuses on whether the encounter history of Sun-like stars in the cluster may be the prohibitive factor in deciding whether the Solar System could form and/or survive. 
Here, we simultaneously combine the two approaches and determine  how many G-dwarfs are within 0.1 -- 0.3\,pc of the supernova 
\citep[in order for the disc to have the correct levels of enrichment,][]{Chevalier00,Looney06,Ouellette07} but do not suffer disruptive dynamical interactions during the course of the cluster's evolution.

In this paper, we use $N$-body simulations to model the evolution of star clusters that contain exactly one  25\,M$_\odot$ star in a cluster with $\sim$2000 other stars and determine 
the number of G-dwarfs that experience the necessary supernova enrichment, where the 25\,M$_\odot$ star is expected to go supernova at $\sim 6.6$\,Myr \citep[e.g.][]{Hirschi04}. Of those G-dwarfs, we also determine their interaction history within the cluster. We model four different initial 
cluster set-ups to cover a large range of potential star formation scenarios, but keep the stellar population constant so that stochastic differences in the clusters' evolution can be identified. 

The paper is organised as follows: in 
Section~\ref{typical} we describe our method of sampling the IMF to obtain a `typical' cluster, in Section~\ref{method} we describe the $N$-body simulations, in Section~\ref{results} we 
describe the results for four different types of cluster initial conditions. We focus on the dynamical histories of enriched G-dwarfs from a representative simulation in 
Section~\ref{history}, we provide a discussion in Section~\ref{discuss} and we conclude in Section~\ref{conclude}.
  
\section{A `typical' cluster}
\label{typical}

There are two distinct methods for populating a (model) star cluster with stars from an IMF; random sampling versus sorted sampling. In the first scenario, the mass of the cloud from which 
stars form is the only upper limit for stellar masses -- for example, in very rare scenarios a 100\,M$_\odot$ cluster could (mathematically) produce a 100\,M$_\odot$ star 
\citep[e.g.][]{Elmegreen06,Parker07}. In the second senario, there is a direct physical dependence between the cluster mass and the most massive star that can form \citep{Weidner06,Weidner13}. 
\citet{Weidner13} claim that the latter scenario is supported by the observation that many clusters follow a relation that is consistent with sorted sampling \citep[though see][]{Maschberger08}. Such a relation would not be fundamental if 
massive stars could be definitively shown to form in (relative) isolation, and recent work by \citet{Lamb10}, \citet{Bressert12} and \citet{Oey13} have shown many tens of O-type stars to be apparently isolated.

However, the issue of random versus sorted sampling is still the subject of much debate, with \citet{Cervino13} claiming that it is statistically impossible to prove one scenario over the other.  
The discussion is relevant here because sorted sampling implies a minimum number of stars ($\sim$2000) is required for a cluster to host a 25\,M$_\odot$ star. Any G-dwarf enriched by the supernova could 
then be subject to dynamical interactions in this populous cluster. If a cluster was populated randomly from the IMF, then a 25\,M$_\odot$ star could form with very few companions -- in this scenario the 
low probability of this cluster forming could then outweigh the probability of a G-dwarf not suffering perturbing interactions in a more populous cluster.  

In this paper we will not consider the dynamical histories of clusters with `unusual' IMFs from random sampling. Instead, we use the results of Monte Carlo experiments from \citet{Parker07} who randomly 
sampled a \citet{Kroupa02} IMF to examine the distribution in cluster mass of clusters that contain only one massive star ($>$17.5\,M$_\odot$). The median cluster mass from $10^4$ realisations of a cluster where the most 
massive star is 25\,M$_\odot$ corresponds to $\sim$2100 stars, which is also consistent with values expected from the sorted sampling method advocated by \citet{Weidner06} -- compare Figs.~4~and~5 in \citet{Parker07}.

\section{Cluster models}
\label{method}

We conduct pure $N$-body simulations of four different dynamical scenarios for cluster evolution, characterised by the initial virial ratio $\alpha_{\rm vir} = T/|\Omega|$, where $T$ and $|\Omega|$ are the 
total kinetic energy and total potential energy of the stars, respectively. We will first discuss clusters in virial equlibrium ($\alpha_{\rm vir} = 0.5$) with a smooth radial profile. We will then discuss substructured  
clusters with three different initial virial states; subvirial (`cool' -- $\alpha_{\rm vir} = 0.3$), virial (`tepid' -- $\alpha_{\rm vir} = 0.5$) and supervirial (`warm' -- $\alpha_{\rm vir} = 0.7$). 

To obtain an idea of the stochasticity in the simulations, we model 20 different realisations of the same cluster. We retain the same stellar population, but set the positions 
and velocities of the stellar systems with a different random number each time. 

\subsection{Dynamical evolution}

We evolve the clusters for 10\,Myr using the \texttt{kira} integrator in the Starlab package \citep[e.g.][]{Zwart99,Zwart01}. This follows the clusters until they have started to dissolve 
and hence contribute to the Galactic field population. \citet{Parker12a} found that at 10\,Myr a significant fraction (20 -- 40\,per cent) of stars are unbound in similar clusters to those modelled here, 
and \citet{Allison10} noted that the more substructured a cluster is, the more likely it is to evaporate on timescales less than 10\,Myr. We do not impose an external Galactic tidal field on the clusters, as this will 
have only a minimal effect in the first 10\,Myr. We  implement  stellar and binary evolution by using the \texttt{SeBa} code \citep{Zwart96,Zwart12}, also within Starlab, which updates 
the evolutionary state of stars more frequently than the timestep of the $N$-body integrator. The combination of \texttt{kira} and \texttt{SeBa} enables us to model the clusters as fully collisional systems 
with accurate stellar evolution (including stellar mergers and binary evolution).

\subsection{Smooth clusters in virial equilibrium}

We model smooth clusters in virial equilibrium using a Plummer sphere \citep{Plummer11}, according to the prescription in \citet*{Aarseth74}. We force the most massive star in the cluster 
to be at the cluster centre, as mass segregation is observed in several large clusters, and smooth, virialised clusters cannot mass segregate dynamically on short timescales \citep{Bonnell98}. 
The Plummer spheres have a half-mass radius of 0.4\,pc.

\subsection{Substructured clusters}

We set up substructured clusters using the fractal prescription in \citet{Goodwin04a}. This has the advantage that the substructure is described by just one parameter, the fractal dimension $D$. 
In three dimensions, a highly substructured cluster has a fractal dimension $D = 1.6$, and a uniformly spherical cluster has $D = 3.0$. We set up clusters with a moderate level of substructure, 
with $D = 2.0$. 

The velocities of the stellar systems (be they single or binary) are correlated according to the substructure; stars that are close have similar velocities, whereas distant stars can have very different 
velocities \citep{Goodwin04a}. We refer the reader to \citet{Goodwin04a,Allison10,Parker11c} for a fuller description of this cluster set-up method. The fractals have a radius of 1\,pc.

We then vary the initial virial ratio of the stars and scale the velocities of the individual stars to the desired virial ratio. In one suite of simulations the clusters are subvirial ($\alpha_{\rm vir} = 0.3$), 
which results in cool collapse during the first 1\,Myr \citep{Allison10}. In this set-up, the most massive stars are placed at random in the fractal -- they may subsequently mass-segregate so that 
the massive stars sink to the cente of the cluster. Another suite of simulations are initially in virial equilibirum ($\alpha_{\rm vir} = 0.5$) -- the initial substructure is subsequently erased through 
dynamical interactions, but the cluster is not expected to form a central core which is as dense as in the cool-collapse clusters. Finally, we run a suite of simulations where the stars are initially 
supervirial ($\alpha_{\rm vir} = 0.7$) -- to determine whether supernova enrichment of G-dwarfs can occur if the birth cluster is globally unbound.

\subsection{Stellar systems}

We place the same population of stellar systems in each cluster to establish that any differences in the evolution of the clusters is due to the random differences in system velocity or position, 
rather than total mass or different binary properties. 

The majority of G-type stars in the Galactic field have a binary companion \citep{Duquennoy91,Raghavan10}. We include binaries in our simulations for the reason that a Sun-like star that experiences the 
necessary amount of supernova enrichment may not be a suitable Solar System analogue if it is in a close ($< 100$\,au) binary system. 

We set the most massive star in the cluster to be 25\,M$_\odot$ and then draw the remaining \emph{primary} masses randomly from a \citet{Kroupa02} IMF of the form
\begin{equation}
 \frac{dN}{dM}   \propto  \left\{ \begin{array}{ll} M^{-1.3} \hspace{0.4cm} m_0
  < M/{\rm M_\odot} \leq m_1   \,, \\ M^{-2.3} \hspace{0.4cm} m_1 <
  M/{\rm M_\odot} \leq m_2   \,,
\end{array} \right.
\end{equation}
where $m_0$ = 0.1\,M$_\odot$, $m_1$ = 0.5\,M$_\odot$, and  $m_2$ = 20\,M$_\odot$, so that we do not have any other $>$20\,M$_\odot$ (O-type) stars in the clusters.

\subsubsection{Binary systems}

We set stellar systems up with the binary fraction and orbital parameters observed in the Galactic field. Note that the field is probably a dynamically evolved population; the primordial binary fraction 
was likely higher, and the period and eccentricity distributions will also have evolved. In principle, it is possible to `reverse engineer' the initial binary population by comparing the observed binary properties in young clusters 
with simulated clusters \citep[e.g.][]{Parker11c,King12a,Geller13}. However, for the purposes of this paper, we simply wish to impose a lower limit on the number of G-dwarfs that reside in binary systems initially and the field population is a suitable lower-limit; we will 
discuss this assumption in Section~\ref{discuss}.

The field binary fraction decreases as a function of primary mass.  Primary masses in the range 0.1~$\leq M/{\rm M}_\odot~<$~0.47 are M-dwarfs, with a 
binary fraction of 0.42 \citep{Fischer92}. K-dwarfs have masses in  the range 0.47~$\leq~M/{\rm M}_\odot$~$<$~0.84 with a binary fraction of 0.45 \citep{Mayor92}, and G-dwarfs have masses 
from 0.84~$\leq~M/{\rm M}_\odot~<$~1.2  with a binary fraction of 0.57 \citep{Duquennoy91,Raghavan10}. All stars more massive than  1.2\,M$_\odot$ are grouped together and assigned a 
binary fraction of unity, as massive stars have a much larger binary fraction than low-mass stars 
\citep[e.g.][and references therein]{Abt90,Mason98,Kouwenhoven05,Kouwenhoven07,Pfalzner07,Mason09}. If a random number exceeds the binary fraction of the primary mass, a secondary 
mass is drawn from a flat mass ratio distribution \citep{Reggiani11a}.

The periods of binary systems in the field are observed to have a log-normal distribution \citep{Duquennoy91,Raghavan10} of the form
\begin{equation}
f\left({\rm log_{10}}P\right)  \propto {\rm exp}\left \{ \frac{-{({\rm log_{10}}P -
\overline{{\rm log_{10}}P})}^2}{2\sigma^2_{{\rm log_{10}}P}}\right \},
\end{equation}
where $\overline{{\rm log_{10}}P} = 4.8$, $\sigma_{{\rm log_{10}}P} = 2.3$ and $P$ is  in days. 

The eccentricities of binary stars are drawn from a thermal distribution \citep{Heggie75} of the form
\begin{equation}
f_e(e) = 2e.
\end{equation}
In the sample of \citet{Duquennoy91}, close binaries (with periods less than 10 days) are almost exclusively on tidally circularised orbits. We account for this by reselecting the eccentricity 
of a system if it exceeds the following period-dependent value :
\begin{equation}
e_{\rm tid} = \frac{1}{2}\left[0.95 + {\rm tanh}\left(0.6\,{\rm log_{10}}P - 1.7\right)\right].
\end{equation}

We combine the primary and secondary masses of the binaries with their semi-major axes and eccentricities to determine the relative velocity and radial components of the stars in each system. \\

We continue this procedure until the cluster has 2100 stars, which corresponds the median cluster mass from $10^4$ realisations of a cluster where the most massive star is 25\,M$_\odot$ \citep{Parker07}. 
The next nine most massive stars in this cluster range from 7 -- 13\,M$_\odot$, and the cluster contains a total of 96 G-dwarfs. The single stars and binaries are then placed randomly at a system position in the fractal or Plummer sphere.

\section{Cluster evolution}
\label{results}

In this Section, we consider four different sets of initial conditions for star cluster formation and follow the subsequent dynamical evolution for 10\,Myr. For each cluster, we determine the 
number of G-dwarfs, $N_{\rm enrich}$, that are within 0.1 -- 0.3\,pc of the the supernova and therefore experience the required levels of isotope enrichment observed in Solar System meteorites 
\citep{Chevalier00,Ouellette07}. Of these $N_{\rm enrich}$ G-dwarfs, we then determine how many are either in close ($<$100\,au) binary systems, or suffer a close ($<$100\,au) encounter that could 
affect the outer regions of the Solar System \citep{Adams01,Adams06,Dukes12}. We label the number of these enriched `singletons' \citep{Malmberg07b} $N_{\rm enrich, sing}$. Finally, we might expect that 
a large velocity perturbation could disrupt planet formation (or a young system of planets). We therefore count the number of enriched singletons that do not suffer a velocity kick greater than 1km\,s$^{-1}$ 
(the typical velocity dispersion in a bound embedded cluster) as being dynamically unperturbed, $N_{\rm enrich, sing, unp}$.

In Section~\ref{history} we will show the dynamical histories of several enriched G-dwarfs in a representative simulation, but in this Section we focus on whether the numbers of enriched G-dwarfs depend on 
the different adopted initial conditions for star formation. 

\subsection{Smooth, virial clusters}

\begin{figure*}
  \begin{center}
\setlength{\subfigcapskip}{10pt}
\hspace*{-0.3cm}
\subfigure[Morphology, 0\,Myr]{\label{plummer_ic}\rotatebox{270}{\includegraphics[scale=0.33]{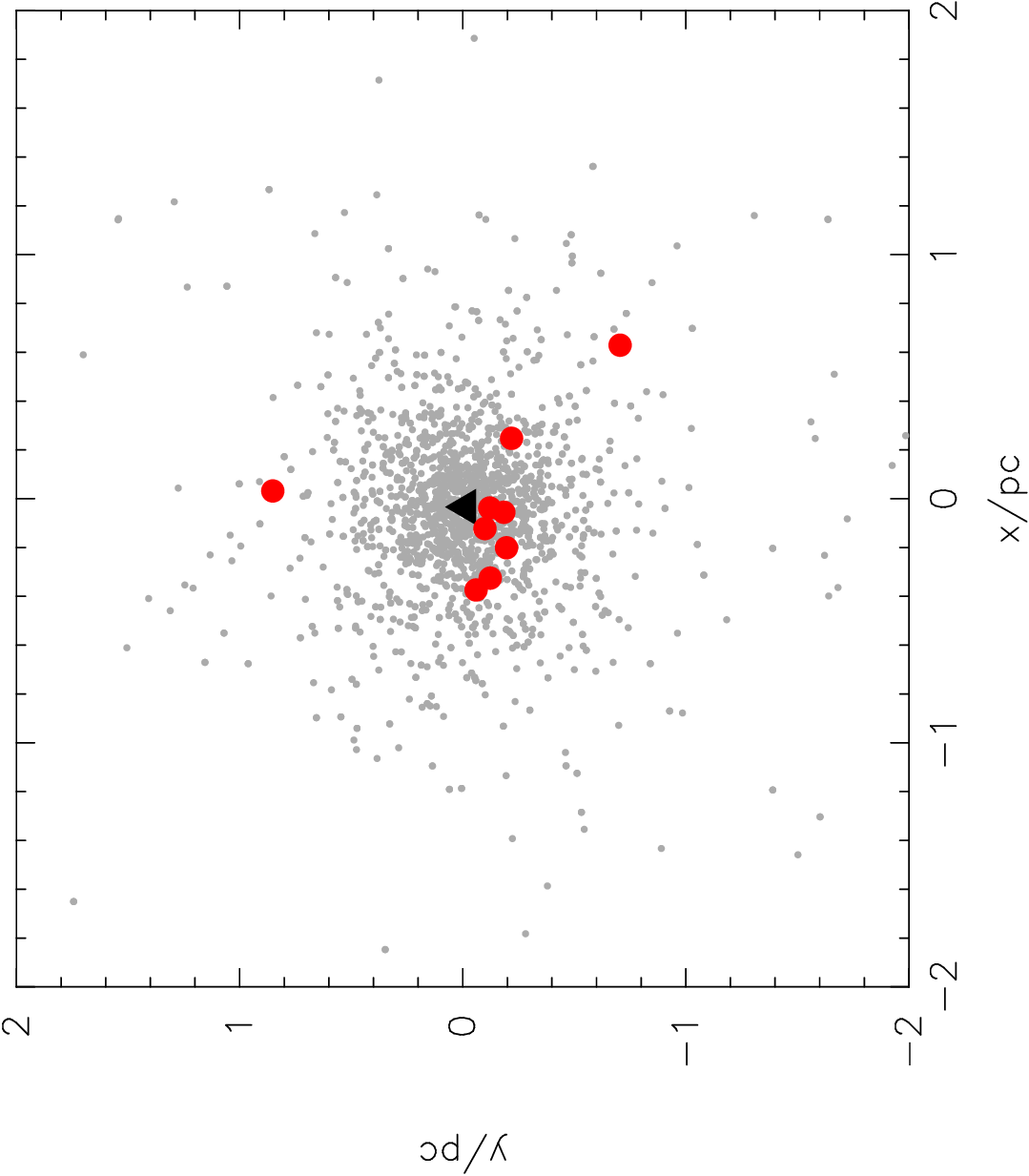}}}  
\hspace*{0.2cm}
\subfigure[Morphology, 6.63\,Myr]{\label{plummer_snapshot}\rotatebox{270}{\includegraphics[scale=0.33]{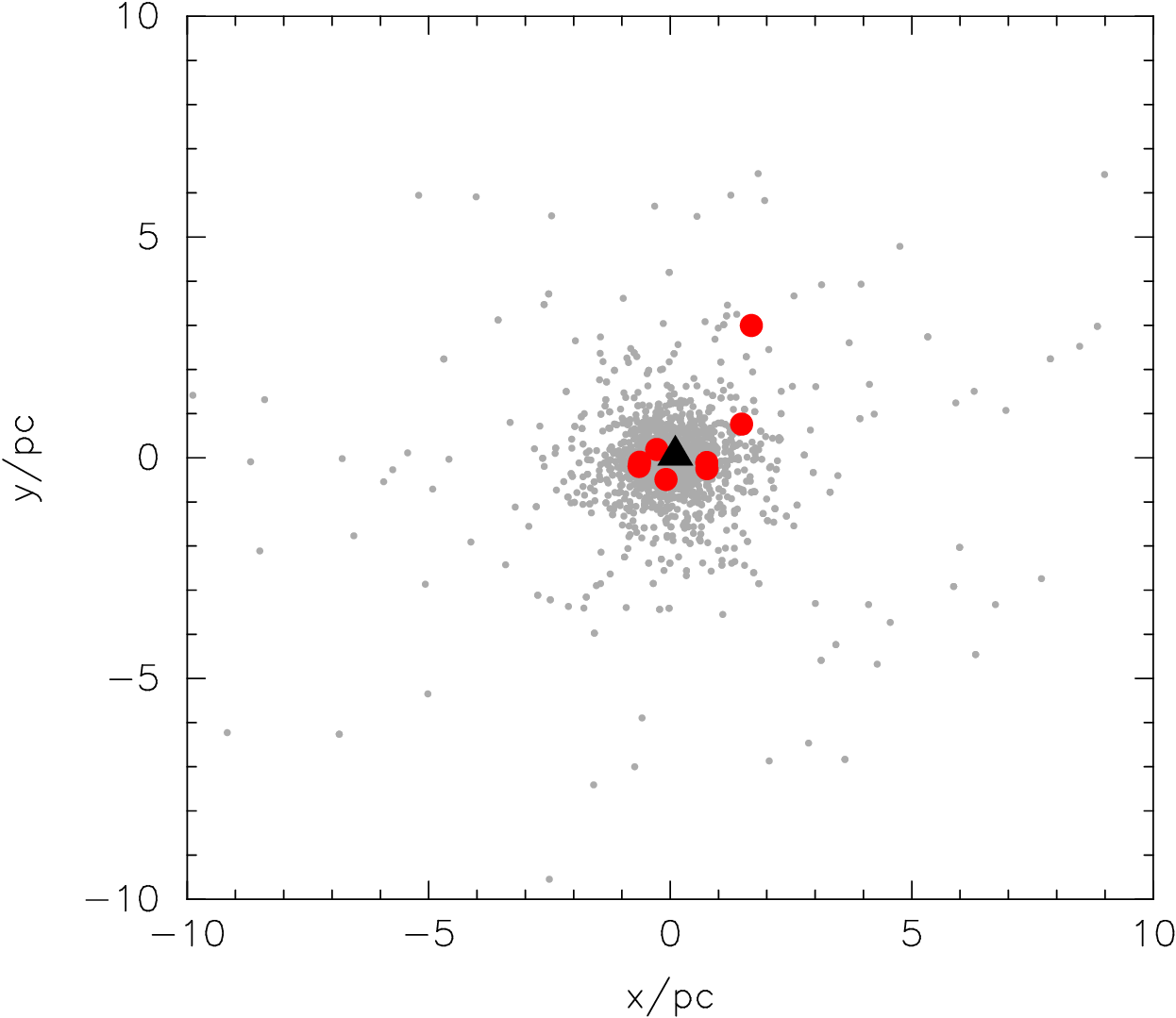}}}
\hspace*{-0.3cm}
\vspace*{0.1cm}\subfigure[Distance from supernova]{\label{plummer_SN_dist}\rotatebox{270}{\includegraphics[scale=0.33]{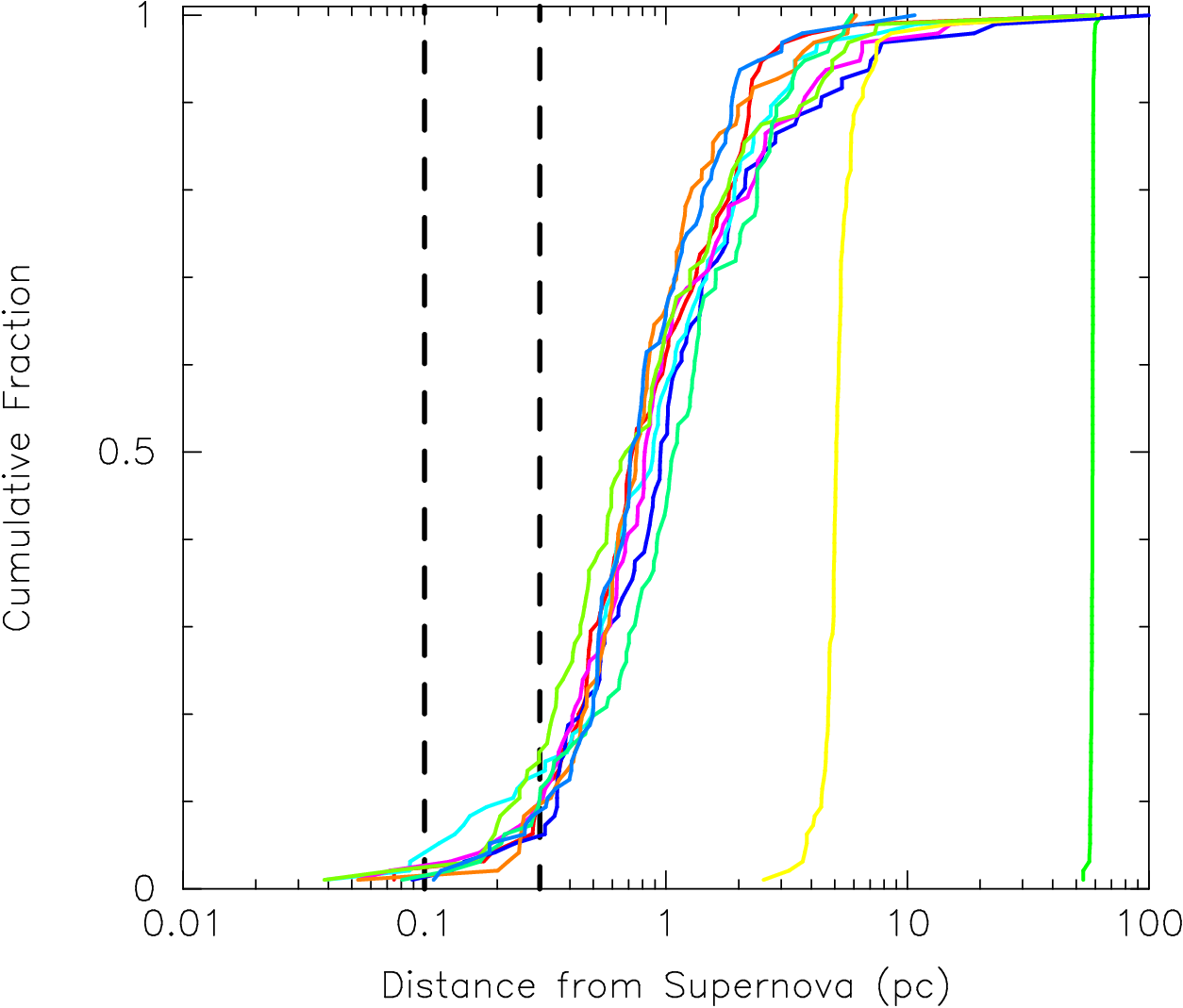}}}
\hspace*{0.2cm}
\subfigure[Numbers of enriched G-dwarfs for all simulations]{\label{plummer_enrich}\rotatebox{270}{\includegraphics[scale=0.33]{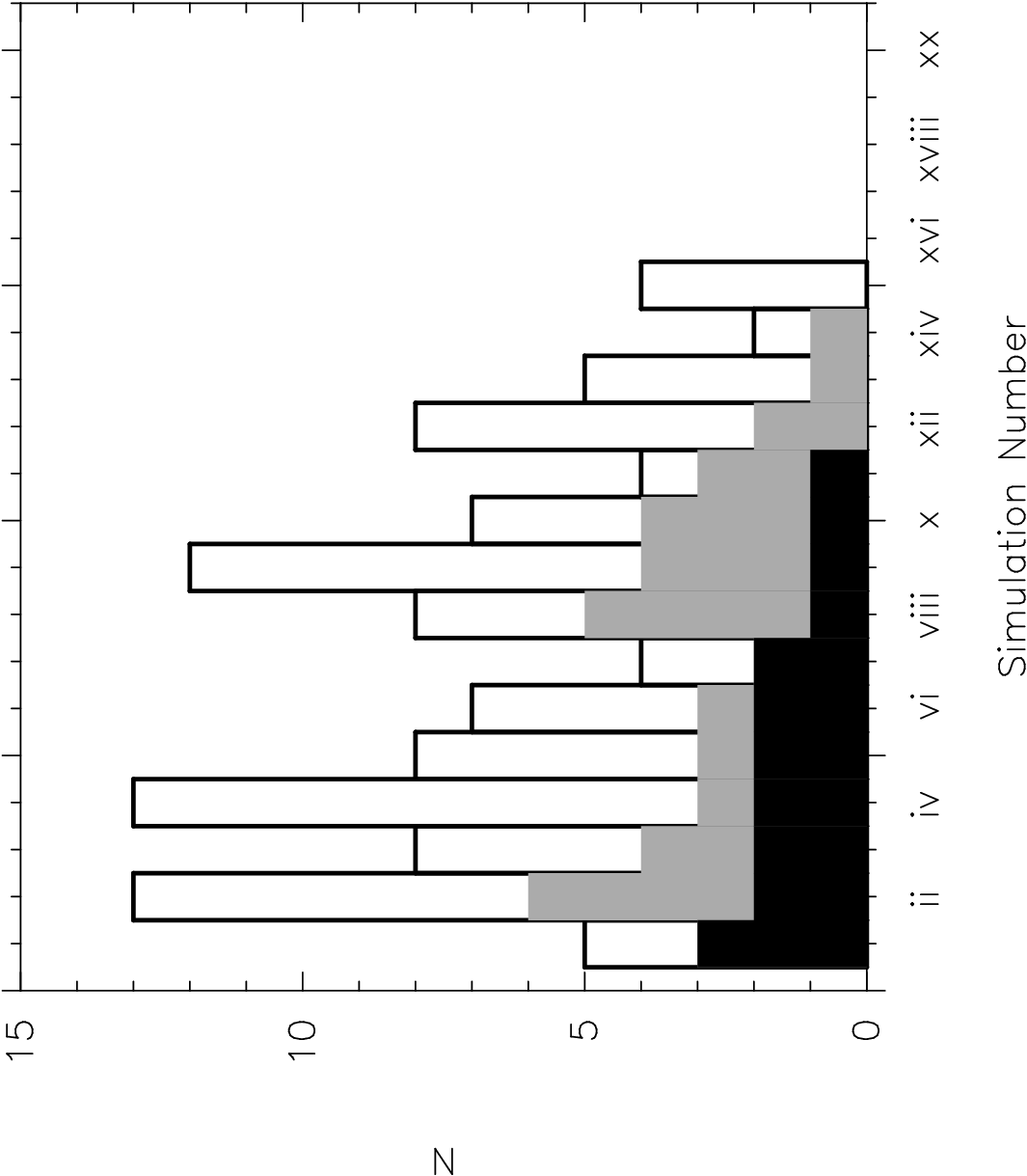}}}
 \end{center}
  \caption[bf]{Dynamical evolution and enrichment in clusters initially in virial equilibrium with a smooth radial profile. In panel (a) we show the initial cluster morphology and  
in panel (b) we show the cluster morphology when the 25\,M$_\odot$ star goes supernova (at 6.63\,Myr) for a typical simulation  (cluster number (i) in Table~\ref{Plummer_data}). The 
position of the supernova progenitor is shown by the black triangle, and the next nine massive stars (with masses 7 -- 13\,M$_\odot$) are shown by the the red filled circles. 
In panel (c) we show the cumulative distributions of G-dwarf distances from the supernova explosion; the 0.1 -- 0.3\,pc `Goldilocks zone' for enrichment of our own Solar 
System is between the two vertical dashed lines. For clarity, we only show the distributions for ten randomly chosen clusters in Table~\ref{Plummer_data}. In panel (d) we show the 
distributions of $N_{\rm enrich}$ (the open histogram), $N_{\rm enrich, sing}$ (the grey histogram) and $N_{\rm enrich, sing, unp}$ (the black histogram) for all twenty simulations.}
  \label{plummer_cluster}
\end{figure*}

We show the typical morphology of a smooth Plummer-sphere cluster at 0\,Myr (Fig.~\ref{plummer_ic}) and at the supernova time (6.63\,Myr -- Fig.~\ref{plummer_snapshot}). The black triangle indicates the position of the supernova progenitor at each time. 
The cumulative distributions of distances from the supernova for all 96 G-dwarfs for 10 clusters is shown in Fig.~\ref{plummer_SN_dist}. The 0.1 -- 0.3\,pc `Goldilocks zone' for enrichment of our own Solar System is between the two vertical dashed lines.

As one might expect, the smooth Plummer-sphere clusters in virial equilibrium follow very similar evolutionary patterns. They all retain a smooth, centrally concentrated morphology. However, the number of enriched G-dwarfs $N_{\rm enrich}$ does vary between clusters, 
as does the number of enriched, unperturbed singletons $N_{\rm enrich, sing, unp}$. In Fig.~\ref{plummer_enrich} we show the distribution of $N_{\rm enrich}$ in all twenty simulations by the open histogram, the distribution of $N_{\rm enrich, sing}$ by the grey histogram and the distribution of $N_{\rm enrich, sing, unp}$ by the black histogram.
Note that the simulations are sorted by $N_{\rm enrich, sing, unp}$, then $N_{\rm enrich, sing}$ and then $N_{\rm enrich}$.

Firstly, 5/20 clusters do not contain any enriched G-dwarfs. This is mainly due to the supernova progenitor interacting with the other massive stars in the cluster and being ejected; however, in one cluster (number (xx) in Table~\ref{Plummer_data}) the 25\,M$_\odot$ star merges with a 5\,M$_\odot$ star and the product 
does not explode as a supernova before the end of the simulation.

In the 15 simulations where we do have supernova enrichment, the number of enriched G-dwarfs, $N_{\rm enrich}$, varies between 2 and 13 (from a total of 96 G-dwarfs). 14/20 clusters contain between 1 and 6 enriched stars that are singletons and 11/20 clusters contain 1, 2, or 3 enriched singletons 
that are unperturbed. We summarise the results in Table~\ref{Plummer_data}.

\begin{table}
\caption[bf]{Data from the simulations in which the cluster is in virial equilibrium with a smooth (Plummer sphere) morphology. We show the simulation number, total number of enriched G-dwarfs, $N_{\rm enrich}$, the number of enriched G-dwarfs that are `singletons' -- i.e.\,\,they are never in a binary or suffer an interaction 
with a semi-major axis less than 100\,au, $N_{\rm enrich, sing}$, and the number of enriched singletons that do not suffer significant velocity perturbations,  $N_{\rm enrich, sing, unp}$.}
\begin{center}
\begin{tabular}{|l|c|c|c|c|}
\hline 
sim.\,\,no. & $N_{\rm enrich}$ &
$N_{\rm enrich, sing}$ & $N_{\rm enrich, sing, unp}$ & \\
\hline 
(i) & 5 & 3 & 3 & \\
(ii) &  13 & 6 & 2 &  \\
(iii) & 8 & 4 & 2 & \\
(iv) & 13 & 3 & 2 & \\
(v) &  8 & 3 & 2 &  \\
(vi) &  7 & 3 & 2 & \\
(vii) & 4 & 2 & 2 & \\
(viii) &  8 & 5 & 1 & \\
(ix) & 12 & 4 & 1 & \\
(x) &  7 & 4 & 1 & \\
(xi) & 4 & 3 & 1 & \\
(xii) & 8 & 2 & 0 & \\
(xiii) & 5 & 1 & 0 & \\
(xiv) & 2 & 1 & 0 & \\
(xv) & 4 & 0 & 0 & \\
(xvi)$^a$ & 0 & 0 & 0 &\\
(xvii)$^a$ &  0 & 0 & 0 & \\
(xviii)$^b$ & 0 & 0 & 0 &  \\
(xix)$^b$ & 0 & 0 & 0 & \\
(xx)$^c$ & 0 & 0 & 0 & \\
\hline
\end{tabular}
\end{center}
\medskip
{$^a$Supernova progenitor ejected after interaction with massive binary.}\newline
{$^b$Supernova progenitor ejected after interactions with other massive stars.}\newline
{$^c$No supernova: 25\,M$_\odot$ star merged with 5\,M$_\odot$ star at 5.73\,Myr.}
\label{Plummer_data}
\end{table}

\subsection{Substructured, subvirial clusters}

\begin{figure*}
  \begin{center}
\setlength{\subfigcapskip}{10pt}
\hspace*{-0.3cm}
\subfigure[Morphology, 0\,Myr]{\label{cool_frac_ic}\rotatebox{270}{\includegraphics[scale=0.33]{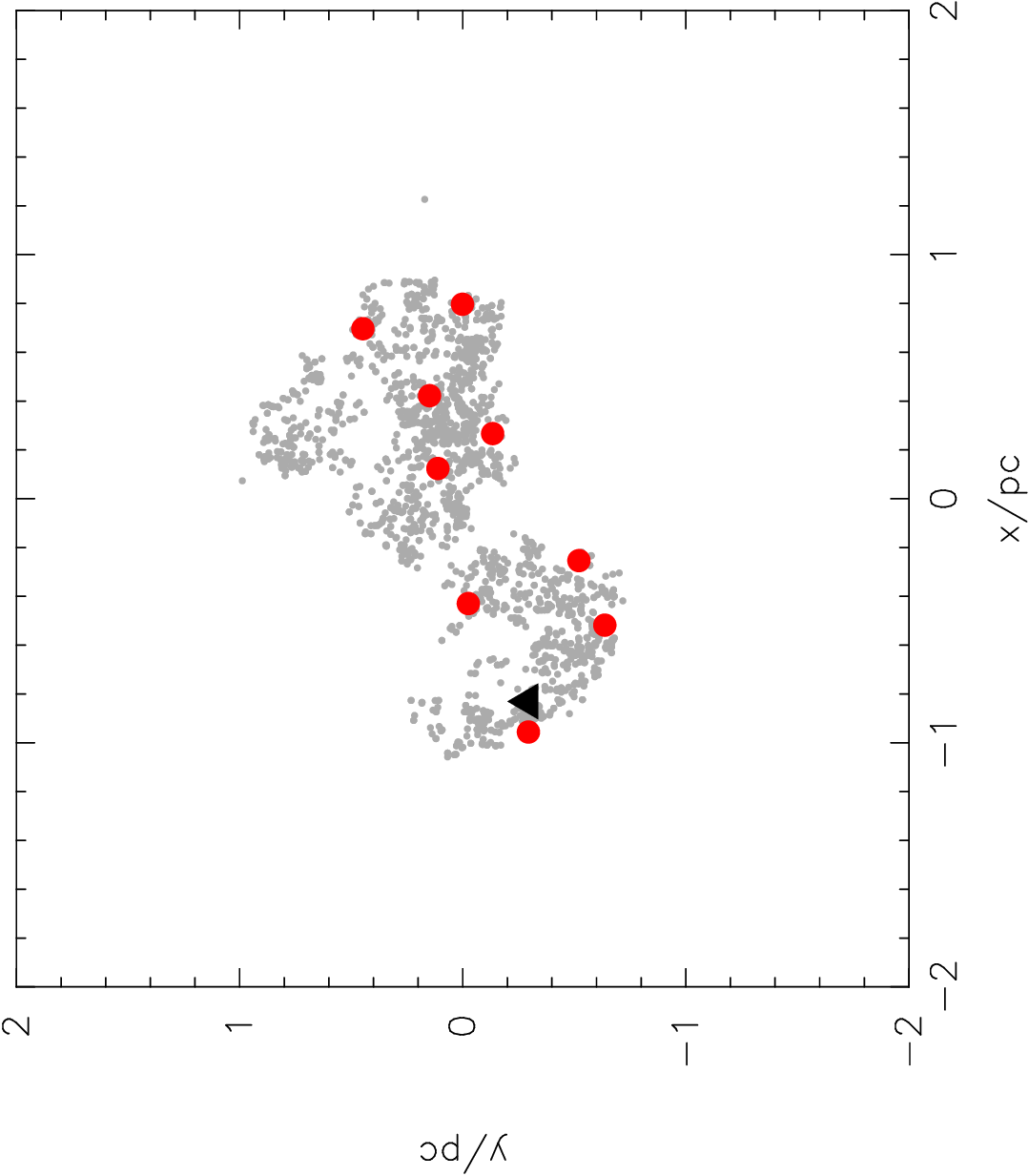}}}  
\hspace*{0.2cm}
\subfigure[Morphology, 6.63\,Myr]{\label{cool_frac_snapshot}\rotatebox{270}{\includegraphics[scale=0.33]{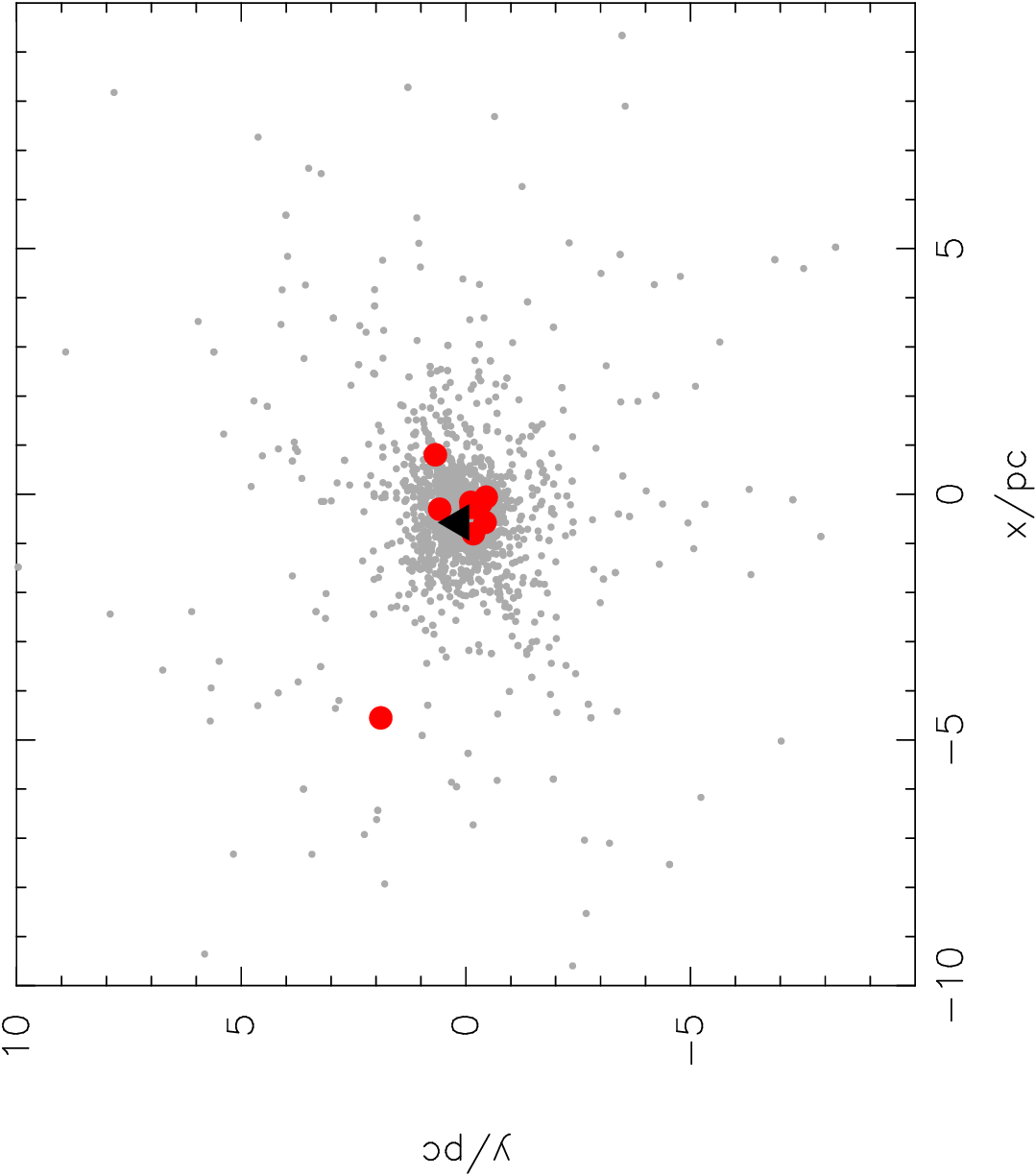}}}

\hspace*{-0.3cm}
\vspace*{0.1cm}\subfigure[Distance from supernova]{\label{cool_frac_SN_dist}\rotatebox{270}{\includegraphics[scale=0.33]{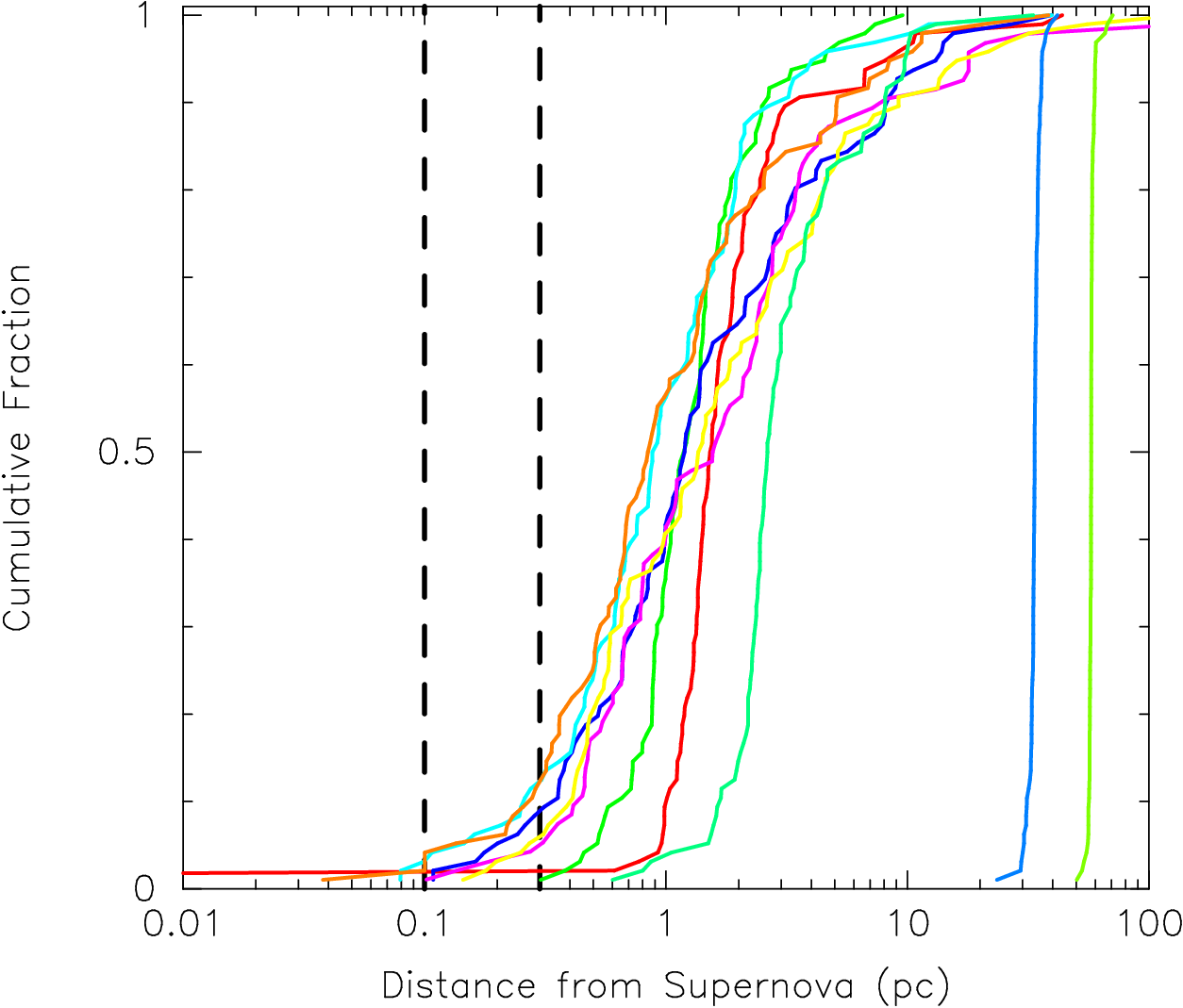}}}
\hspace*{0.2cm}
\subfigure[Numbers of enriched G-dwarfs for all simulations]{\label{cool_frac_enrich}\rotatebox{270}{\includegraphics[scale=0.33]{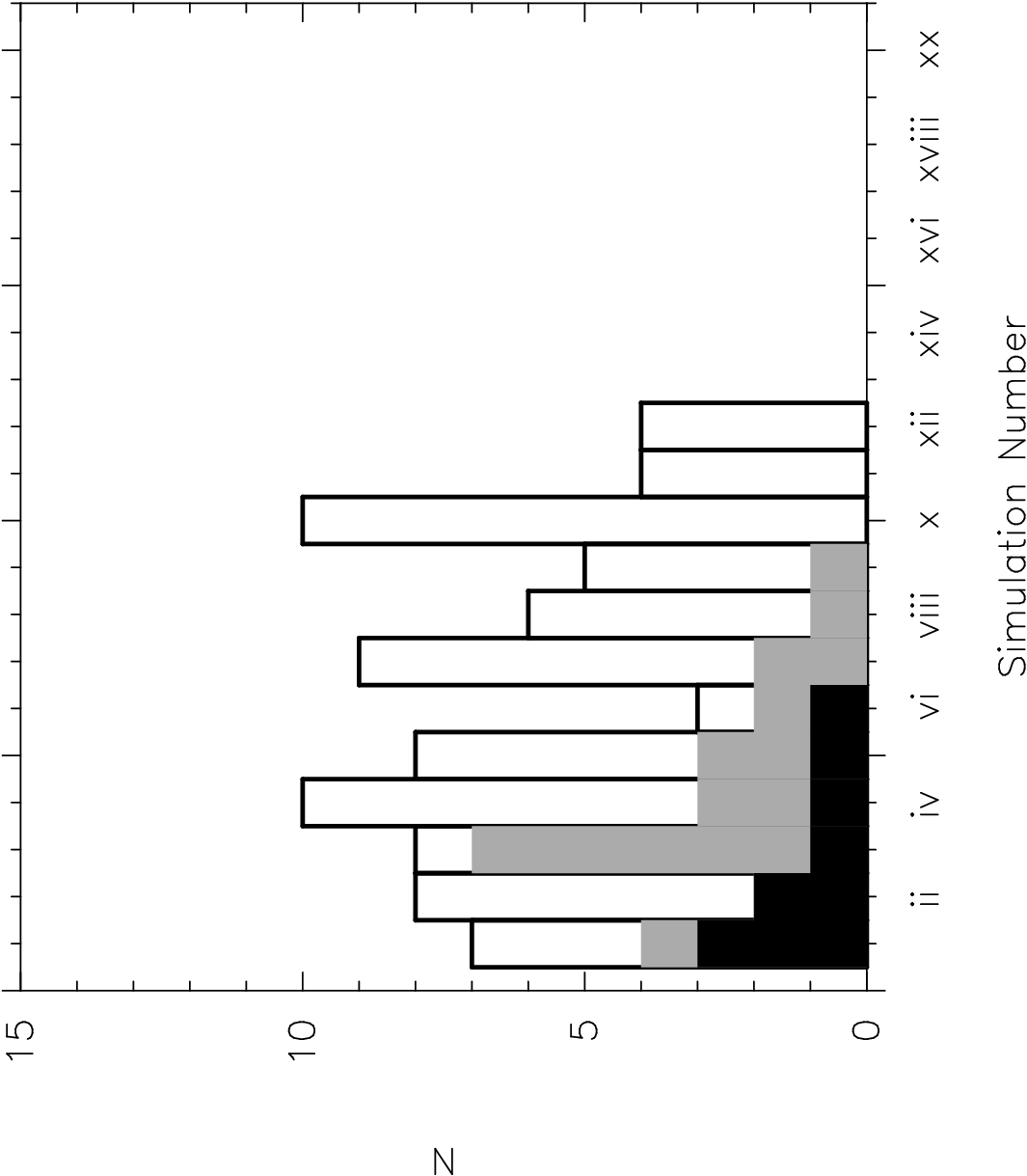}}}
 \end{center}
  \caption[bf]{Dynamical evolution and enrichment in initially substructured, subvirial (cool) clusters. In panel (a) we show the initial cluster morphology and  
in panel (b) we show the cluster morphology when the 25\,M$_\odot$ star goes supernova (at 6.63\,Myr) for a typical simulation (cluster number (vi) in Table~\ref{cool_fractal_data}). 
The position of the supernova progenitor is shown by the black triangle, and the next nine massive stars (with masses 7 -- 13\,M$_\odot$) are shown by the the red filled circles. 
In panel (c) we show the cumulative distributions of G-dwarf distances from the supernova explosion; the 0.1 -- 0.3\,pc `Goldilocks zone' for enrichment of our own Solar System is between the two vertical dashed lines. 
For clarity, we only show the distributions for ten randomly chosen clusters in Table~\ref{cool_fractal_data}. In panel (d) we show the 
distributions of $N_{\rm enrich}$ (the open histogram), $N_{\rm enrich, sing}$ (the grey histogram) and $N_{\rm enrich, sing, unp}$ (the black histogram) for all twenty simulations.}
  \label{cool_frac_cluster}
\end{figure*}

We show the typical morphology of a substructured, subvirial cluster at 0\,Myr (Fig.~\ref{cool_frac_ic}) and at the supernova time (6.63\,Myr -- Fig.~\ref{cool_frac_snapshot}). The black triangle indicates the position of the supernova progenitor each time. We see that 
the initial substructure has been completely erased, and the cluster has now assumed a smooth, centrally concentrated profile with a dense core. The most massive stars have dynamically mass segregated and are at the centre of the cluster, implying that the G-dwarfs that are enriched by 
the supernova must pass through the centre of the cluster at least once.

\begin{table}
\caption[bf]{Data from the simulations in which the cluster is substructured and subvirial. We show the simulation number, the total number of enriched G-dwarfs, $N_{\rm enrich}$, the number of enriched G-dwarfs that are `singletons' -- i.e.\,\,they are never in a binary or suffer an interaction 
with a semi-major axis less than 100\,au, $N_{\rm enrich, sing}$, and the number of enriched singletons that do not suffer significant velocity perturbations,  $N_{\rm enrich, sing, unp}$.}
\begin{center}
\begin{tabular}{|l|c|c|c|c|}
\hline 
sim.\,\,no. & $N_{\rm enrich}$ &
$N_{\rm enrich, sing}$ & $N_{\rm enrich, sing, unp}$ &\\
\hline 
(i)  & 7 & 4 & 3 & \\
(ii)  & 8 & 2 & 2 & \\
(iii)  & 8  & 7 & 1 & \\
(iv)  & 10 & 3 & 1 & \\
(v)  & 8 & 3 & 1 & \\
(vi)  & 3 & 2 & 1 & \\
(vii)  & 9 & 2 & 0 & \\
(viii)  & 6 & 1 & 0 & \\
(ix)  & 5 & 1 & 0 & \\
(x)  & 10 & 0 & 0 & \\
(xi)  & 4 & 0 & 0 & \\
(xii)  & 4 & 0 & 0 & \\
(xiii)$^a$  & 0 & 0 & 0 & \\
(xiv)$^b$  & 0 & 0 & 0 & \\
(xv)$^b$  & 0 & 0 & 0 & \\
(xvi)$^c$  & 0 & 0 & 0 & \\
(xvii)$^c$  & 0 & 0 & 0 & \\
(xviii)$^d$  & 0 & 0 & 0 & \\
(xix)$^d$  & 0 & 0 & 0 & \\
(xx)$^d$  & 0 & 0 & 0 & \\
\hline
\end{tabular}
\end{center}
 \medskip
{$^a$Supernova progenitor ejected after interaction with massive binary.}\newline
{$^b$Supernova progenitor ejected after interactions with other massive stars.}\newline
{$^c$Supernova progenitor ejected from unstable Trapezium-like system.}\newline
{$^d$No G-dwarfs between 0.1 -- 0.3\,pc from the supernova due to the cluster's rapid expansion.}
\label{cool_fractal_data}
\end{table}
 
Substructured clusters evolve very stochastically, especially those undergoing cool collapse \citep{Allison10,Parker12b}. The violent relaxation process these clusters undergo can lead to the ejection of massive stars, often after they have become the member of an unstable 
Trapezium-like system \citep{Allison11}. 

The cumulative distributions of distances from the supernova for all 96 G-dwarfs for 10 clusters is shown in Fig.~\ref{cool_frac_SN_dist}. The 0.1 -- 0.3\,pc `Goldilocks zone' for enrichment of our own Solar System is between the two vertical dashed lines.

In Fig.~\ref{cool_frac_enrich} we show the distribution of $N_{\rm enrich}$ in all twenty simulations by the open histogram, the distribution of $N_{\rm enrich, sing}$ by the grey histogram and the distribution of $N_{\rm enrich, sing, unp}$ by the black histogram.

8/20 clusters do not have any enriched G-dwarfs, often due to the supernova progenitor interacting with the other massive stars in short-lived and unstable Trapezium-like systems, resulting in the ejection of the progenitor. Examples of this are shown in Fig.~\ref{cool_frac_SN_dist} by the green 
and blue lines on the far right of the panel, which are the cumulative distributions of G-dwarf distances from simulations (xvi) and (xvii) in  Table~\ref{cool_fractal_data}.

In the 12 simulations where we do have supernova enrichment, the number of enriched G-dwarfs, $N_{\rm enrich}$, varies between 3 and 10 (from a total of 96 G-dwarfs). 9/20 clusters contain between 1 and 7 enriched stars that are singletons and 6/20 clusters contain 1, 2, or 3 enriched singletons 
that are unperturbed, $N_{\rm enrich, sing, unp}$. We summarise the results in Table~\ref{cool_fractal_data}.

\subsection{Substructured, virial clusters}

\begin{figure*}
  \begin{center}
\setlength{\subfigcapskip}{10pt}
\hspace*{-0.3cm}
\subfigure[Morphology, 0\,Myr]{\label{tepid_frac_ic}\rotatebox{270}{\includegraphics[scale=0.33]{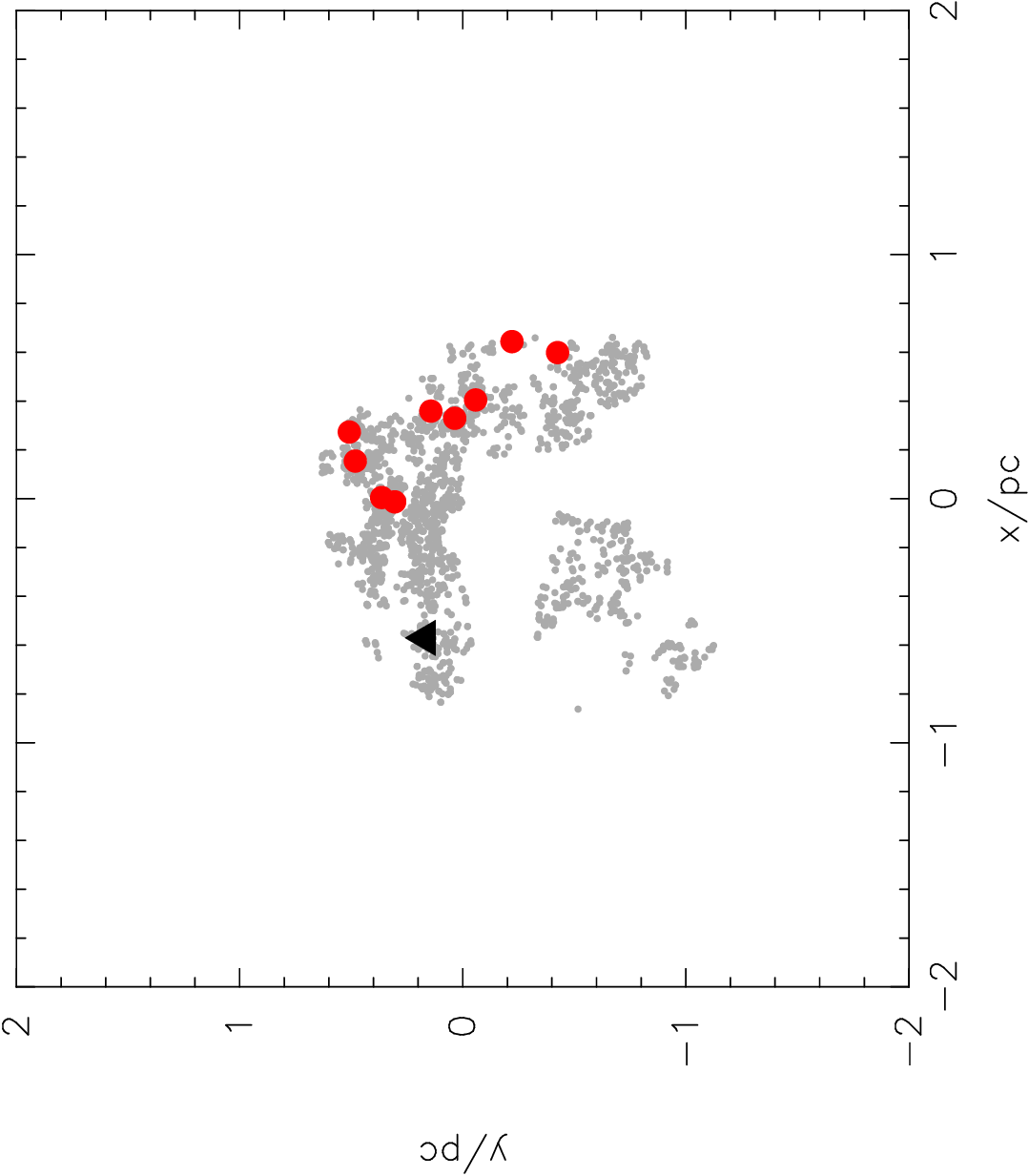}}}  
\hspace*{0.2cm}
\subfigure[Morphology, 6.65\,Myr]{\label{tepid_frac_snapshot}\rotatebox{270}{\includegraphics[scale=0.33]{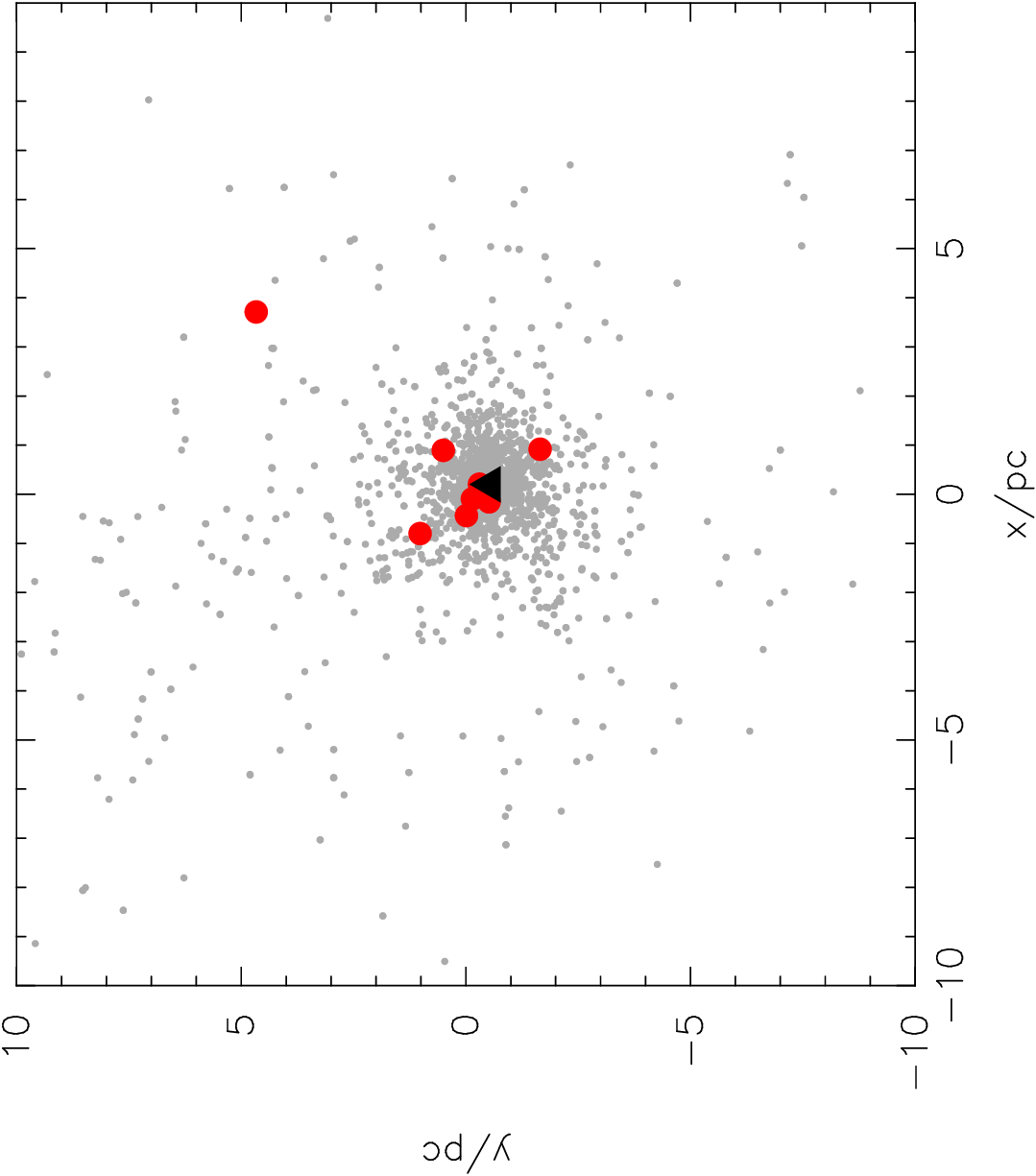}}}
\hspace*{-0.3cm}
\vspace*{0.1cm}\subfigure[Distance from supernova]{\label{tepid_frac_SN_dist}\rotatebox{270}{\includegraphics[scale=0.33]{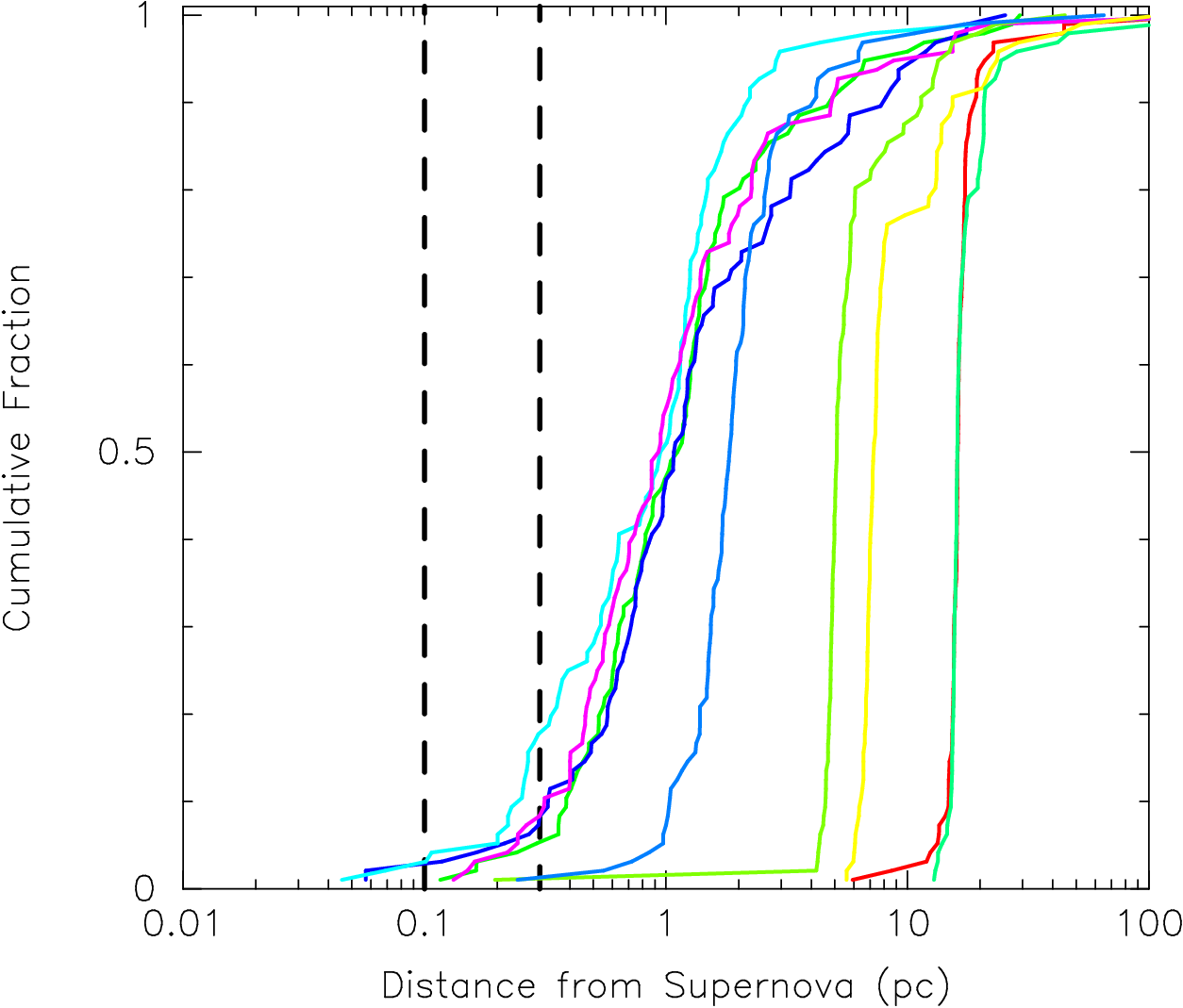}}}
\hspace*{0.2cm}
\subfigure[Numbers of enriched G-dwarfs for all simulations]{\label{tepid_frac_enrich}\rotatebox{270}{\includegraphics[scale=0.33]{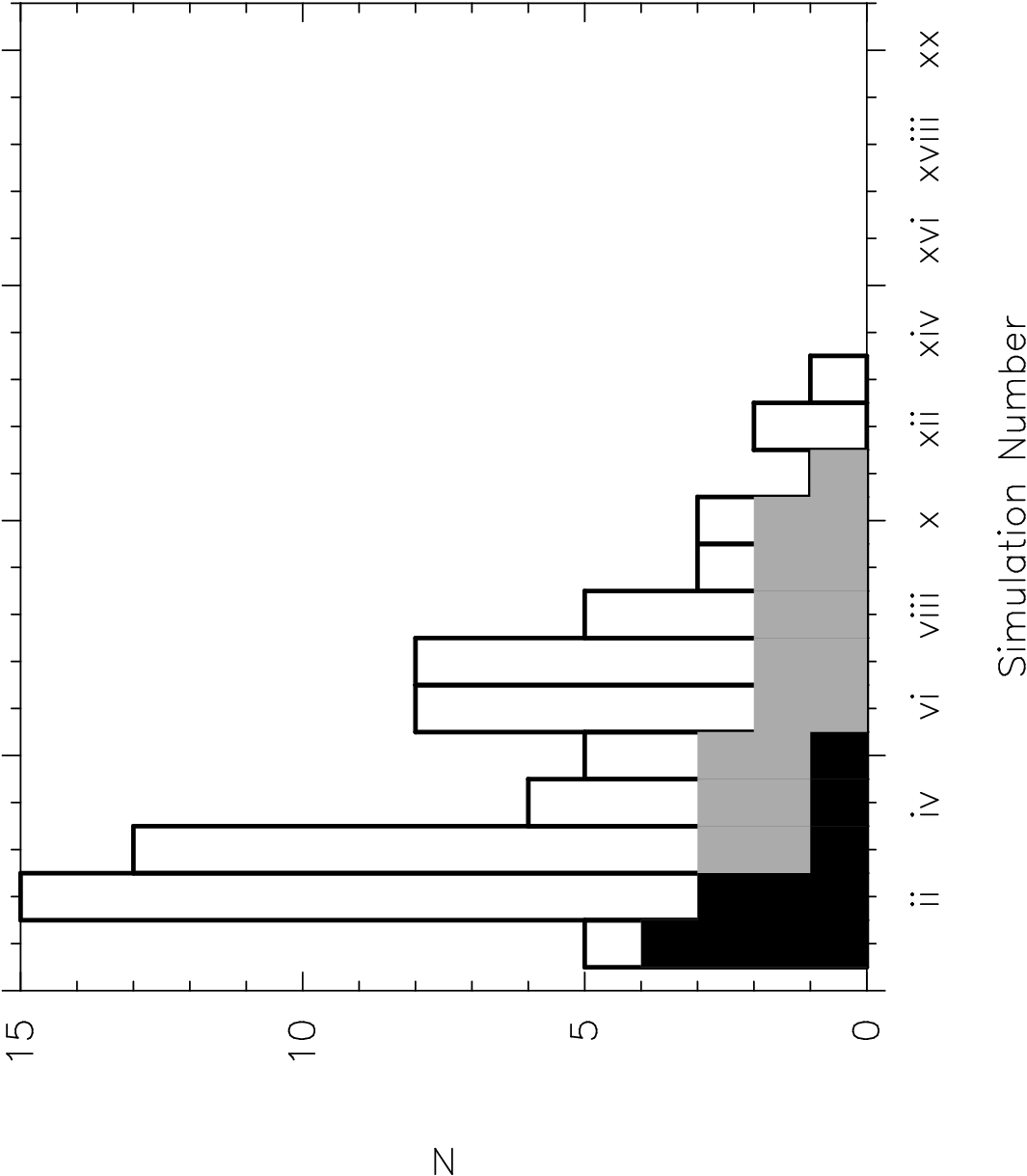}}}
 \end{center}
  \caption[bf]{Dynamical evolution and enrichment in initially substructured, virialised (tepid) clusters. In panel (a) we show the initial cluster morphology and  
in panel (b) we show the cluster morphology when the 25\,M$_\odot$ star goes supernova (at 6.65\,Myr) for a typical simulation (cluster (i) in 
Table~\ref{tepid_fractal_data}). The position of the supernova progenitor is shown by the black triangle, and the next nine massive stars (with masses 7 -- 13\,M$_\odot$) 
are shown by the the red filled circles. In panel (c) we show the cumulative distributions of G-dwarf distances from the supernova explosion; the 0.1 -- 0.3\,pc 
`Goldilocks zone' for enrichment of our own Solar System is between the two vertical dashed lines. For clarity, we only show the distributions for ten randomly chosen clusters in Table~\ref{tepid_fractal_data}. 
In panel (d) we show the distributions of $N_{\rm enrich}$ (the open histogram), $N_{\rm enrich, sing}$ (the grey histogram) and $N_{\rm enrich, sing, unp}$ (the black histogram) for all twenty simulations.}
  \label{tepid_frac_cluster}
\end{figure*}

We show the typical morphology of an initially substructured, virialised cluster at 0\,Myr (Fig.~\ref{tepid_frac_ic}) and at the supernova time (6.65\,Myr -- Fig.~\ref{tepid_frac_snapshot}). The black triangle indicates the position of the supernova progenitor each time. The majority of these clusters 
lose substructure within the first few Myr of evolution, although one cluster (number (xx) in Table~\ref{tepid_fractal_data}) retains substructure in the sense that it forms a `binary cluster' -- a cluster with two distinct groups of stars. 

Substructured, virialised clusters are also dense enough to eject the supernova progenitor -- again, usually from unstable Trapezium-like systems. The binary cluster (xx) is not dense enough to have any G-dwarfs in the 0.1 -- 0.3 pc 
enrichment range, although as we will see in the case of supervirial clusters, the formation of a binary cluster does not preclude enrichment.  

The cumulative distributions of distances from the supernova for all 96 G-dwarfs for 10 clusters is shown in Fig.~\ref{tepid_frac_SN_dist}. The 0.1 -- 0.3\,pc `Goldilocks zone' for enrichment of our own Solar System is between the two vertical dashed lines.

In Fig.~\ref{tepid_frac_enrich} we show the distribution of $N_{\rm enrich}$ in all twenty simulations by the open histogram, the distribution of $N_{\rm enrich, sing}$ by the grey histogram and the distribution of $N_{\rm enrich, sing, unp}$ by the black histogram.

7/20 clusters do not have any enriched G-dwarfs. In the 13 simulations where we do have supernova enrichment, the number of enriched G-dwarfs, $N_{\rm enrich}$, varies between 1 and 13 (from a total of 96 G-dwarfs). 11/20 clusters contain between 1 and 4 enriched stars that are singletons and 
5/20 clusters contain 1, 3, or 4 enriched singletons that are unperturbed. We summarise the results in Table~\ref{tepid_fractal_data}.

\begin{table}
\caption[bf]{Data from the simulations in which the cluster is substructured and in virial equilibrium. We show the simulation number, the total number of enriched G-dwarfs, $N_{\rm enrich}$, the number of enriched G-dwarfs that are `singletons' -- i.e.\,\,they are never in a binary or suffer an interaction 
with a semi-major axis less than 100\,au, $N_{\rm enrich, sing}$, and the number of enriched singletons that do not suffer significant velocity perturbations,  $N_{\rm enrich, sing, unp}$.}
\begin{center}
\begin{tabular}{|l|c|c|c|c|}
\hline 
sim.\,\,no. & $N_{\rm enrich}$ &
$N_{\rm enrich, sing}$ & $N_{\rm enrich, sing, unp}$ & \\
\hline 
(i) & 5 & 4 & 4 & \\
(ii) & 15 & 3 & 3 & \\
(iii) & 13 & 3 & 1 & \\
(iv) & 6 & 3 & 1 & \\
(v) & 5 & 3 & 1 & \\
(vi) & 8 & 2 & 0 & \\
(vii) & 8 & 2 & 0 & \\
(viii) & 5 & 2 & 0 & \\
(ix) & 3 & 2 & 0 & \\
(x) & 3 & 2 & 0 & \\ 
(xi) & 1 & 1 & 0 &  \\
(xii) & 2 & 0 & 0 & \\
(xiii) & 1 & 0 & 0 & \\
(xiv)$^a$ & 0 & 0 & 0 & \\
(xv)$^b$ & 0 & 0 & 0 & \\
(xvi)$^c$ & 0 & 0 & 0 &\\
(xvii)$^c$ & 0 & 0 & 0 & \\
(xviii)$^c$ & 0 & 0 & 0 & \\
(xix)$^c$ & 0 & 0 & 0 &\\
(xx)$^d$ & 0 & 0 & 0 &  \\
\hline
\end{tabular}
\end{center}
\medskip
{$^a$Supernova progenitor ejected after interaction with massive binary.}\newline
{$^b$Supernova progenitor ejected after interactions with other massive stars.}\newline
{$^c$Supernova progenitor ejected from unstable Trapezium-like system.}\newline
{$^d$No G-dwarfs between 0.1 -- 0.3\,pc from the supernova due to fractal evolving into an extended binary cluster.}
\label{tepid_fractal_data}
\end{table}

\subsection{Substructured, supervirial clusters}

\begin{figure*}
  \begin{center}
\setlength{\subfigcapskip}{10pt}
\hspace*{-0.3cm}
\subfigure[Morphology, 0\,Myr]{\label{warm_frac_ic}\rotatebox{270}{\includegraphics[scale=0.33]{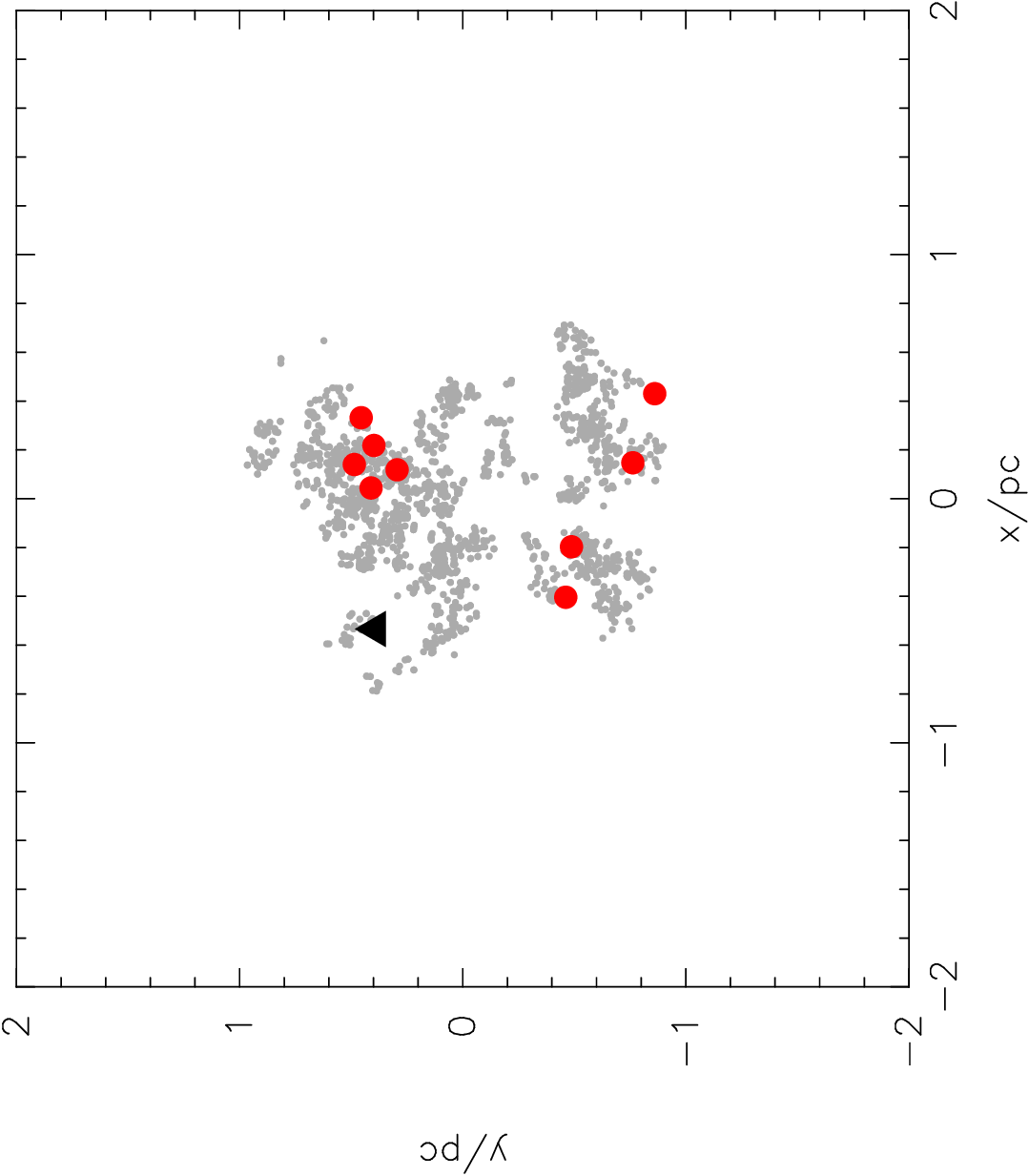}}}  
\hspace*{0.2cm}
\subfigure[Morphology, 6.64\,Myr]{\label{warm_frac_snapshot}\rotatebox{270}{\includegraphics[scale=0.33]{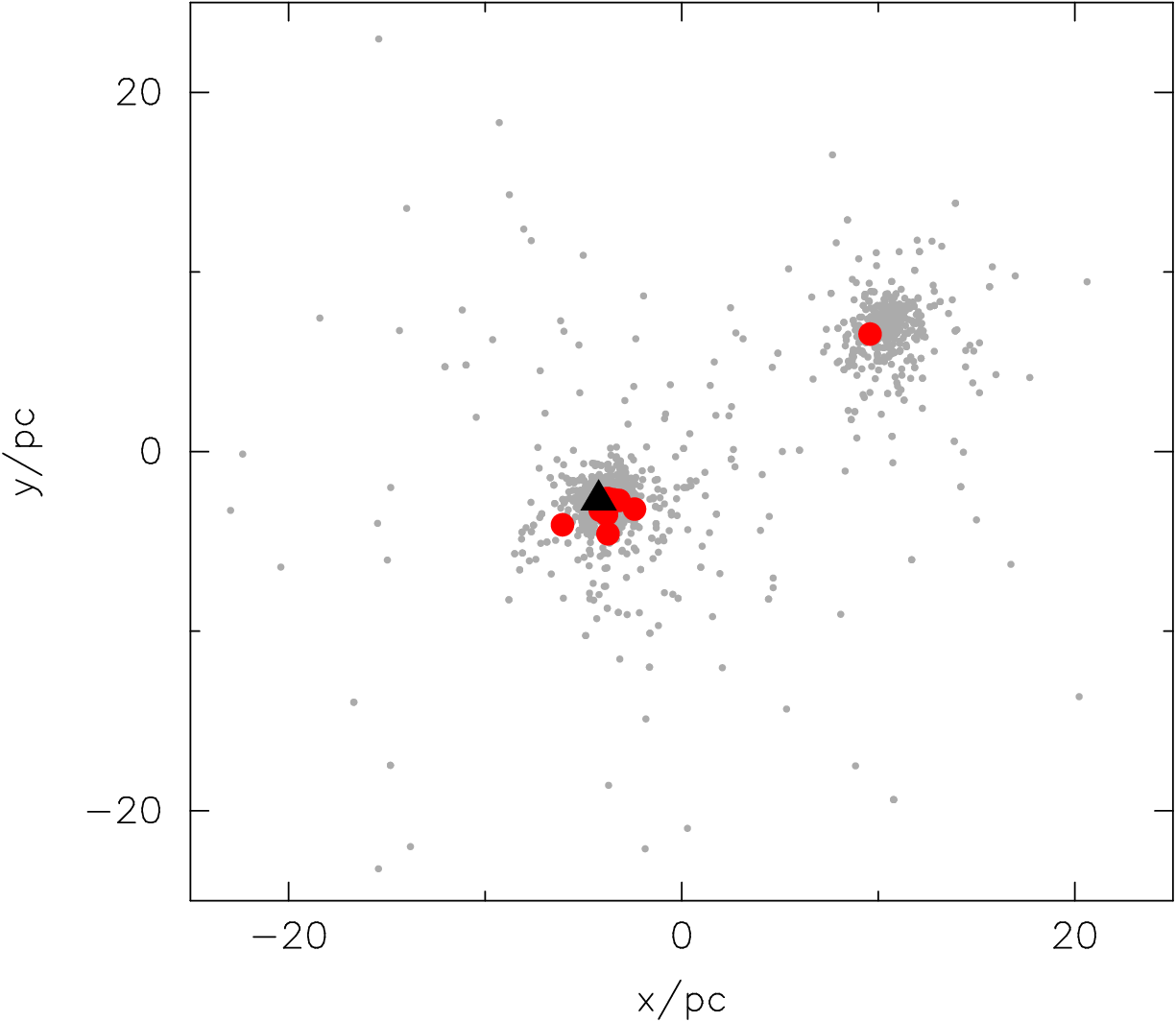}}}
\hspace*{-0.3cm}
\vspace*{0.1cm}\subfigure[Distance from supernova]{\label{warm_frac_SN_dist}\rotatebox{270}{\includegraphics[scale=0.33]{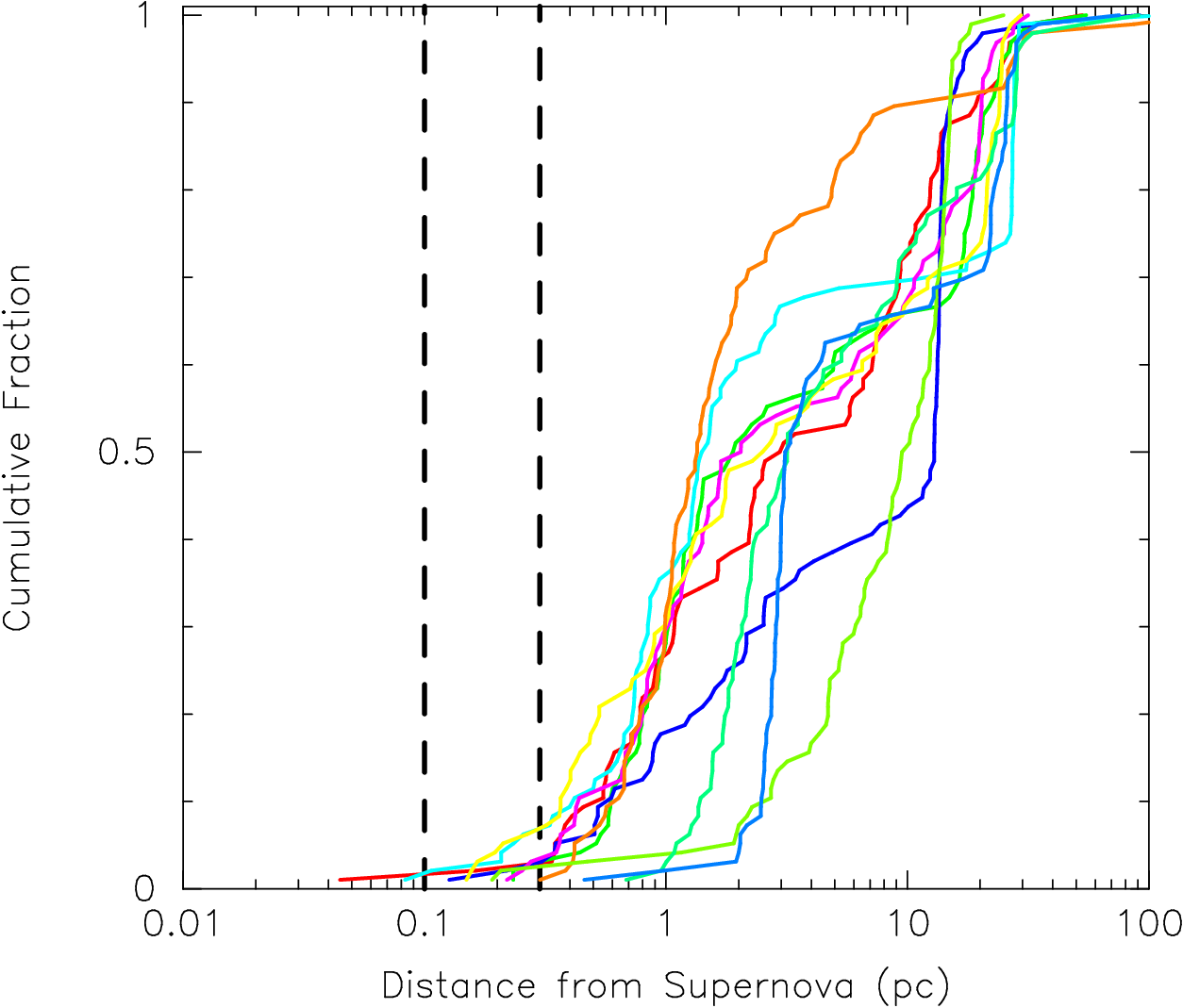}}}
\hspace*{0.2cm}
\subfigure[Numbers of enriched G-dwarfs for all simulations]{\label{warm_frac_enrich}\rotatebox{270}{\includegraphics[scale=0.33]{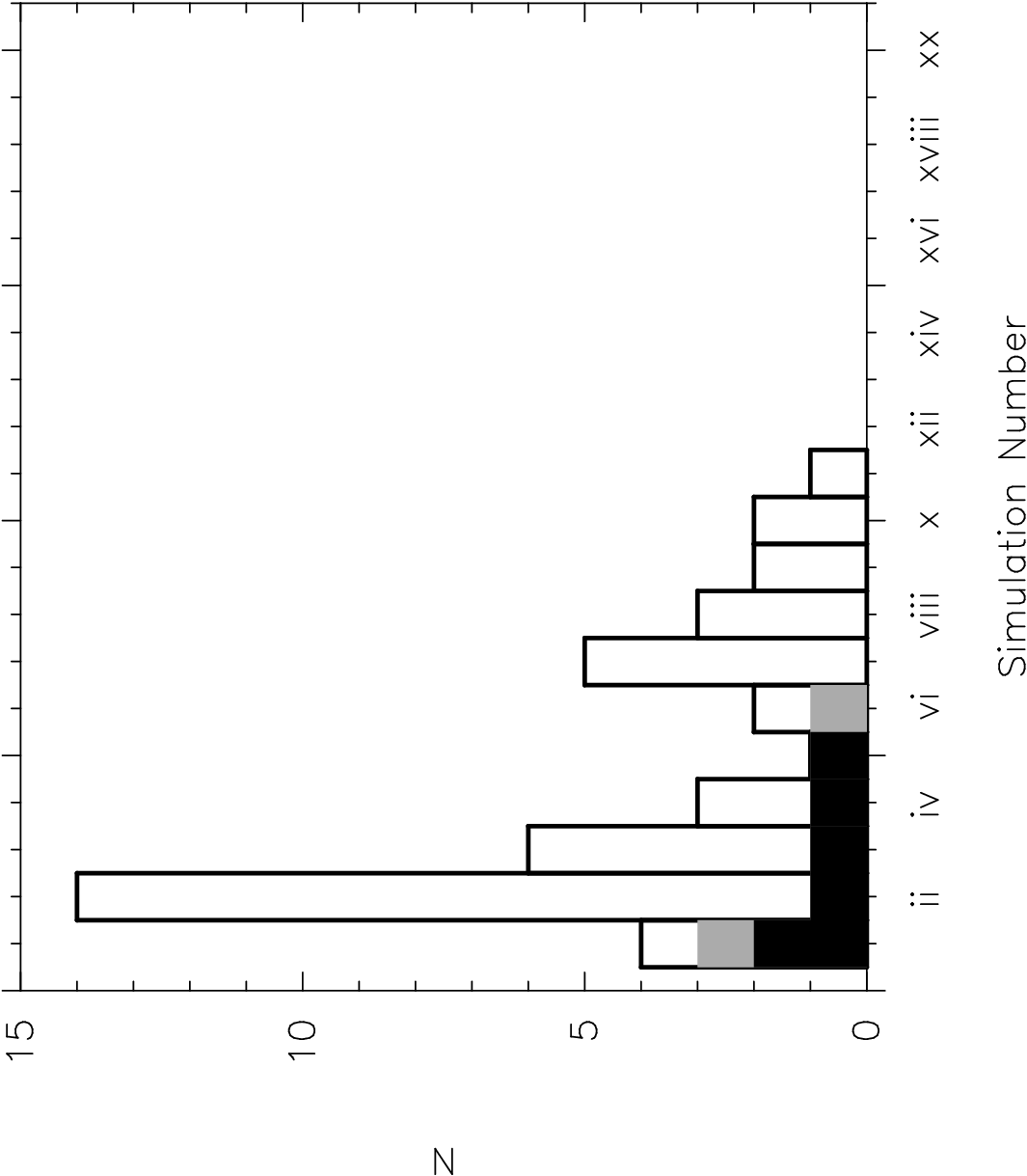}}}
 \end{center}
  \caption[bf]{Dyamical evolution and enrichment for initially substructured, supervirial (warm) clusters. In panel (a) we show the initial cluster morphology and  
in panel (b) we show the cluster morphology when the 25\,M$_\odot$ star goes supernova (at 6.64\,Myr) for a typical simulation (cluster (i) in Table~\ref{warm_fractal_data}). 
The position of the supernova progenitor is shown by the black triangle, and the next nine massive stars (with masses 7 -- 13\,M$_\odot$) are shown by the the red filled circles. 
In panel (c) we show the cumulative distributions of G-dwarf distances from the supernova explosion; the 0.1 -- 0.3\,pc `Goldilocks zone' for enrichment of our own Solar System 
is between the two vertical dashed lines. For clarity, we only show the distributions for ten randomly chosen clusters in Table~\ref{warm_fractal_data}. In panel (d) we show the distributions 
of $N_{\rm enrich}$ (the open histogram), $N_{\rm enrich, sing}$ (the grey histogram) and $N_{\rm enrich, sing, unp}$ (the black histogram) for all twenty simulations.}
  \label{warm_frac_cluster}
\end{figure*}

We show the typical morphology of an initially substructured, supervirial cluster at 0\,Myr (Fig.~\ref{warm_frac_ic}) and at the supernova time (6.64\,Myr -- Fig.~\ref{warm_frac_snapshot}). The black triangle indicates the position of the supernova progenitor each time. 

Substructured, supervirial clusters tend to retain some substructure \citep{Parker12d}. In around half of these simulations, a binary cluster forms (Fig.~\ref{warm_frac_snapshot}) -- with the most massive star (i.e. the supernova progenitor) likely to be in one of the binary `nodes'. Even though the global motion causes the 
cluster to expand, these nodes are dense enough for enrichment to occur in around half of the clusters. The remainder of these supervirial fractals expand to form association--like structures, which are too diffuse for the required G-dwarf enrichment (the clusters with $N_{\rm enrich} = 0$ in Table~\ref{warm_fractal_data}).

The cumulative distributions of distances from the supernova for all 96 G-dwarfs for 10 clusters is shown in Fig.~\ref{warm_frac_SN_dist}. The 0.1 -- 0.3\,pc `Goldilocks zone' for enrichment of our own Solar System is between the two vertical dashed lines. Binary clusters 
are betrayed in this plot by the step-like cumulative distributions in some simulations. 

In Fig.~\ref{warm_frac_enrich} we show the distribution of $N_{\rm enrich}$ in all twenty simulations by the open histogram, the distribution of $N_{\rm enrich, sing}$ by the grey histogram and the distribution of $N_{\rm enrich, sing, unp}$ by the black histogram.

9/20 clusters do not have any enriched G-dwarfs. In the 11 simulations where we do have supernova enrichment, the number of enriched G-dwarfs, $N_{\rm enrich}$, varies between 1 and 14 (from a total of 96 G-dwarfs). 6/20 clusters contain between 1 and 3 enriched stars that are singletons and 5/20 clusters contain 1 or 2 enriched singletons 
that are unperturbed. We summarise the results in Table~\ref{warm_fractal_data}.

\begin{table}
\caption[bf]{Data from the simulations in which the cluster is substructured and supervirial. We show the simulation number, the total number of enriched G-dwarfs, $N_{\rm enrich}$, the number of enriched G-dwarfs that are `singletons' -- i.e.\,\,they are never in a binary or suffer an interaction 
with a semi-major axis less than 100\,au, $N_{\rm enrich, sing}$, and the number of enriched singletons that do not suffer significant velocity perturbations,  $N_{\rm  enrich, sing, unp}$.}
\begin{center}
\begin{tabular}{|l|c|c|c|c|}
\hline 
sim.\,\,no. & $N_{\rm enrich}$ &
$N_{\rm enrich, sing}$ & $N_{\rm enrich, sing, unp}$ & \\
 \hline 
(i) & 4 & 3 & 2 & \\
(ii) & 14 & 1 & 1 & \\
(iii) &  6 & 1 & 1 & \\
(iv) &  3 & 1 & 1 & \\
(v) & 1 & 1 & 1 & \\
(vi) &  2 & 1 & 0 &  \\
(vii) &  5 & 0 & 0 & \\
(viii) &  3 & 0 & 0 & \\
(ix) &  2 & 0 & 0 & \\
(x) & 2 & 0 & 0 & \\ 
(xi) &  1 & 0 & 0 & \\
(xii)$^a$ & 0 & 0 & 0 & \\
(xiii)$^a$ & 0 & 0 & 0 & \\
(xiv)$^a$ & 0 & 0 & 0 & \\
(xv)$^a$ & 0 & 0 & 0 & \\
(xvi)$^a$ & 0 & 0 & 0 & \\
(xvii)$^a$ & 0 & 0 & 0 & \\
(xviii)$^a$ & 0 & 0 & 0 & \\
(xix)$^a$ & 0 & 0 & 0 & \\
(xx)$^a$ & 0 & 0 & 0 & \\
\hline
\end{tabular}
\end{center}
{$^a$No G-dwarfs between 0.1 -- 0.3\,pc from the supernova due to fractal evolving into a diffuse association.}
\label{warm_fractal_data}
\end{table}

\section{Dynamical histories}
\label{history}

In the previous section we showed the number of G-dwarf stars that were enriched at a distance between 0.1 -- 0.3\,pc from the supernova ($N_{\rm enrich}$) as a function of initial cluster conditions. Of those $N_{\rm enrich}$, we determined 
the number $N_{\rm enrich, sing}$ that were singletons, meaning that they were never in a close binary system, nor did they suffer a sub-100\,au encounter. Either scenario would most likely preclude the formation and stable evolution of our Solar System. 
Finally, we applied a stricter criterion that an enriched singleton suffering a strong velocity perturbation 
(i.e.\,\,in excess of the typical velocity dispersion in a cluster) could also be prohibited from forming a stable Solar System. The number of systems that are enriched singletons that do not suffer such perturbations is $N_{\rm enrich, sing, unp}$. 

Here, we focus on the dynamical histories of 3 enriched G-dwarfs in a representative simulation where the cluster is initially substructured and subvirial [number (vi) in Table~\ref{cool_fractal_data}]. Because substructure is erased on short timescales ($<$2\,Myr) in subvirial and virial clusters, the subsequent dynamical 
evolution of the cluster is similar, irrespective of the assumed initial morphology. The only exception is a supervirial (unbound) cluster, which expands and so does not have a well-defined `centre'.  However, when supernova enrichment does occur 
in supervirial clusters, the parameters we concentrate on here (nearest neighbour distance to the enriched G-dwarf, velocity perturbations) are similar for all initial morphologies and virial states.  

\begin{figure*}
  \begin{center}
\setlength{\subfigcapskip}{10pt}
\hspace*{-0.3cm}
\subfigure[Nearest neighbour distance]{\label{fractal_NN_dist}\rotatebox{270}{\includegraphics[scale=0.27]{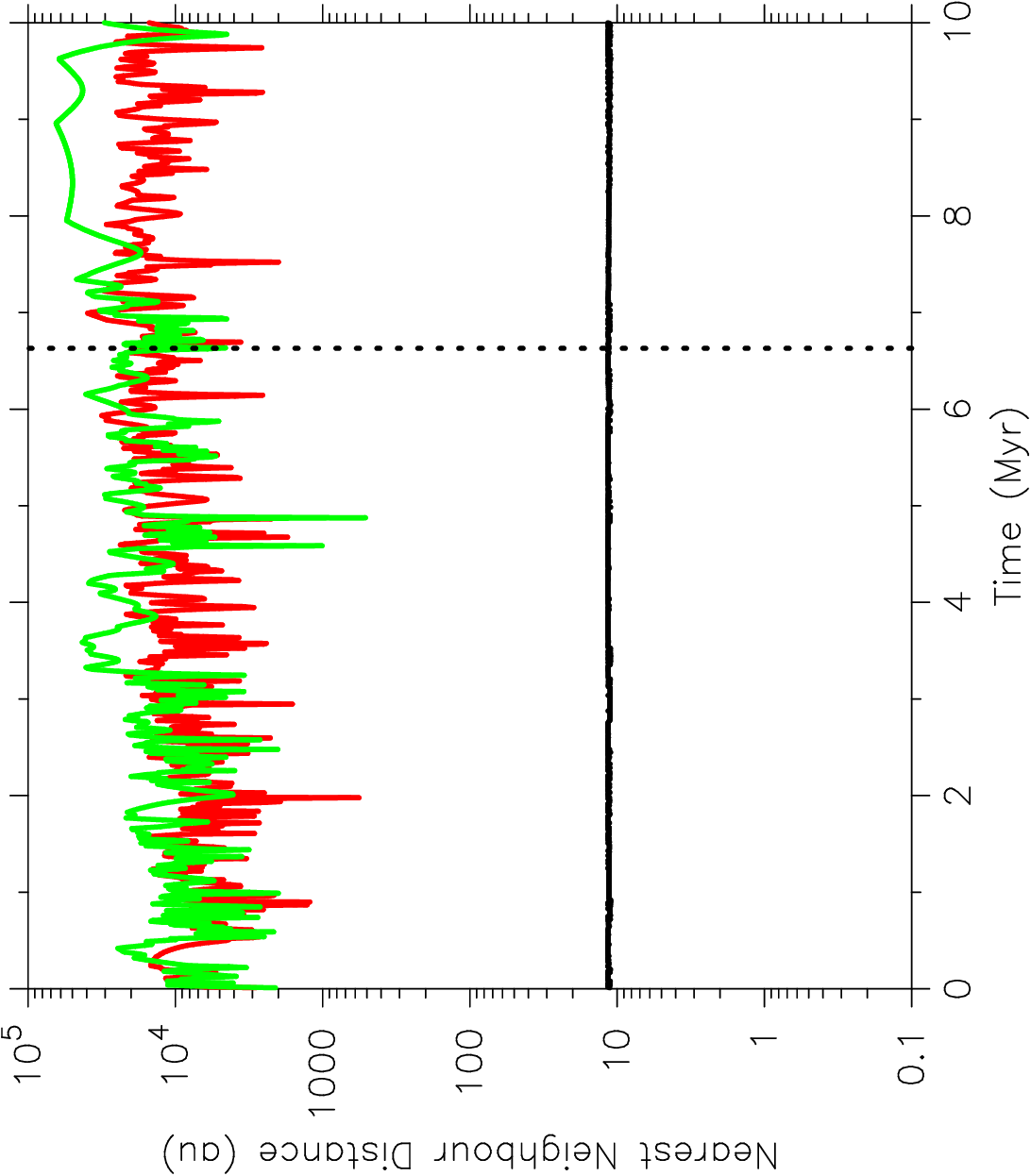}}}
\hspace*{0.2cm}
\subfigure[Distance from cluster centre]{\label{fractal_CC_dist}\rotatebox{270}{\includegraphics[scale=0.27]{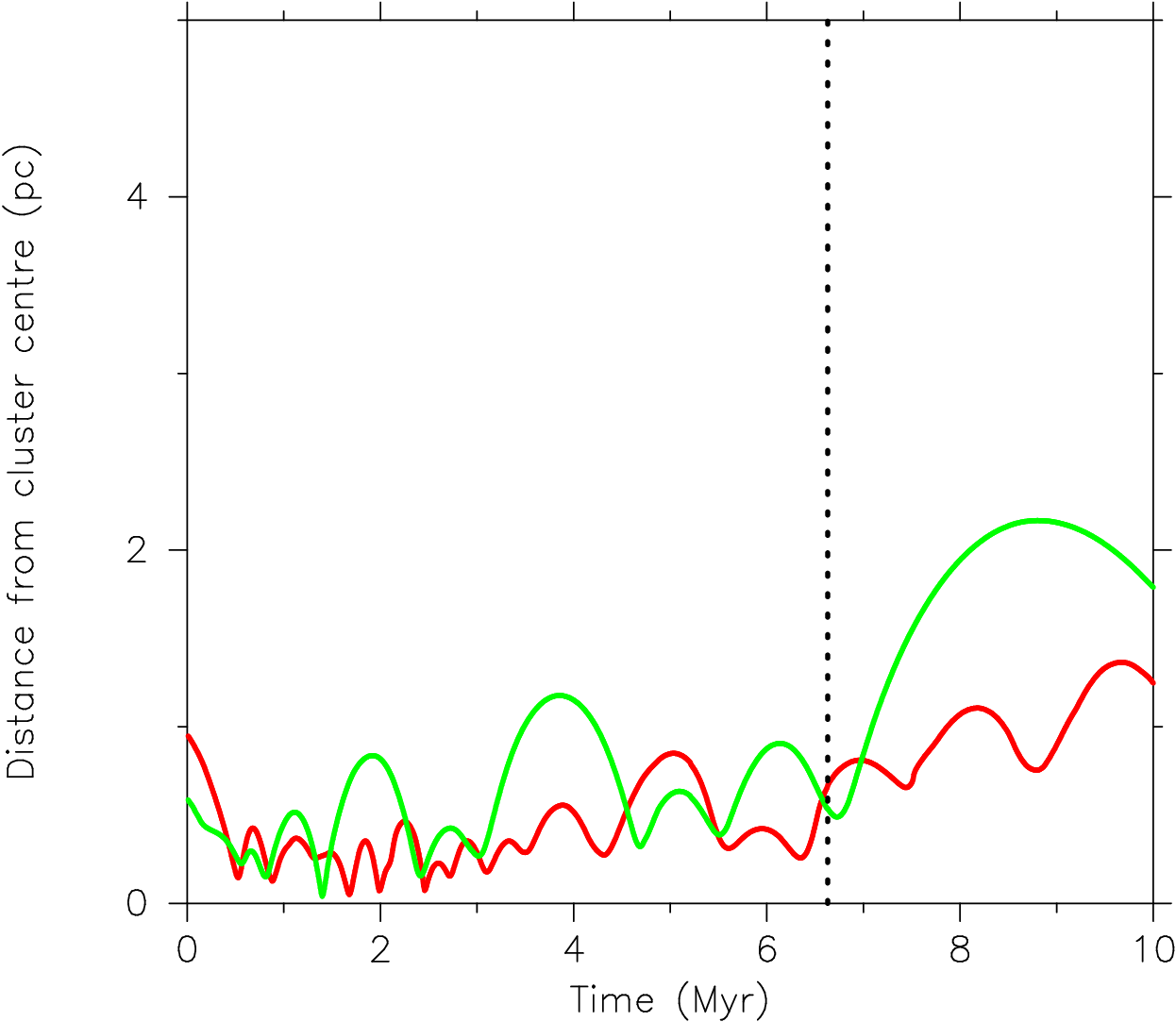}}}
\hspace*{0.2cm}
\subfigure[Change in velocity vector]{\label{fractal_Delta_V}\rotatebox{270}{\includegraphics[scale=0.27]{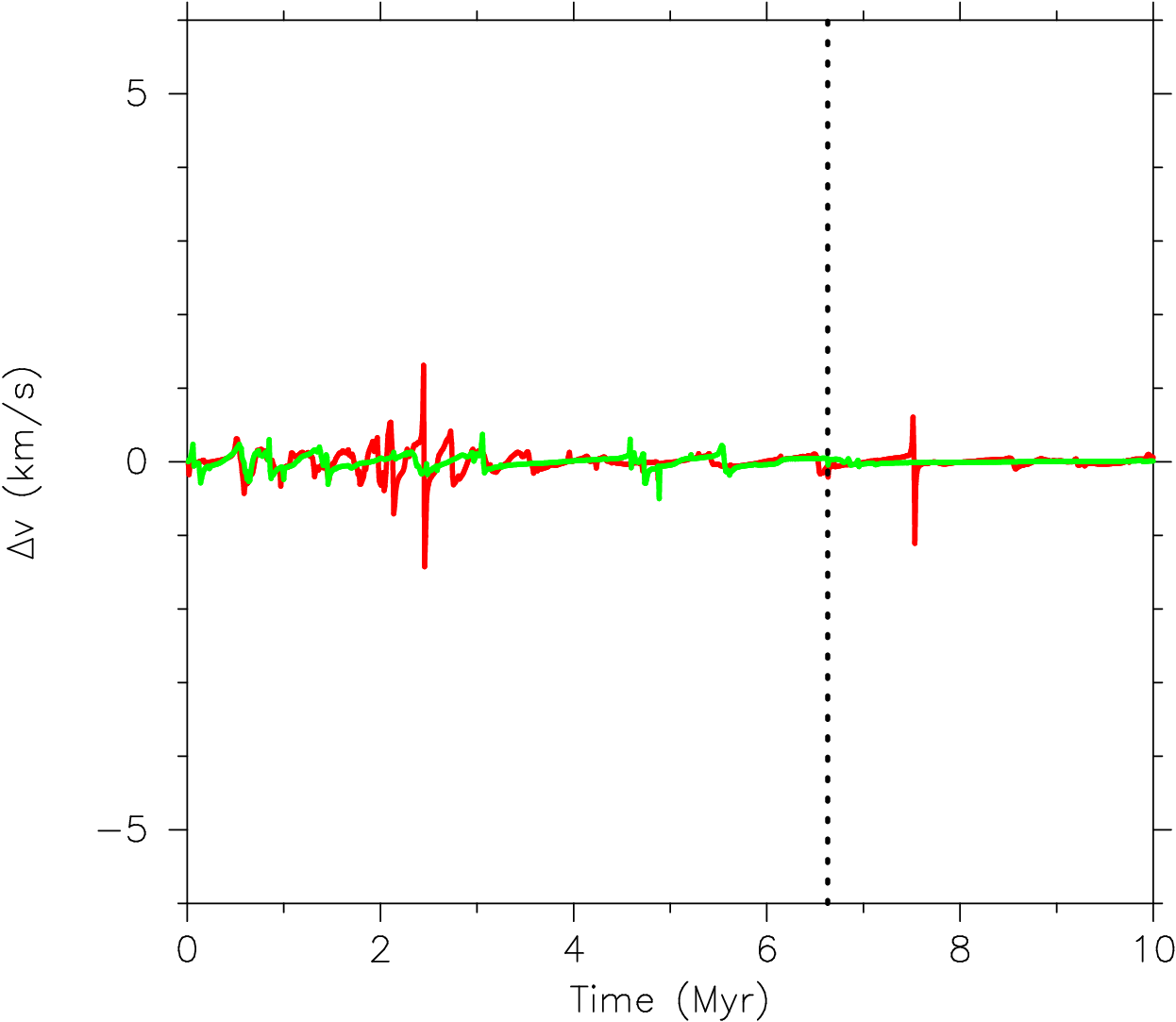}}}
 \end{center}
  \caption[bf]{The dynamical histories of enriched G-dwarfs in a typical simulation run for a substructured, subvirial cluster (number (vi) in Table~\ref{cool_fractal_data}). In panel (a)  we show the nearest neighbour distance; in panel (b)  we show the distance 
from the cluster centre; and in panel (c) we show the change in velocity vector for each of these stars as a function of time. The supernova time is shown by the vertical dotted line. Of the three stars that 
are enriched, two are singletons (shown by the red and green lines), but one suffers at least one significant velocity kick ($>$1\,km\,s$^{-1}$) during the first 10\,Myr of the cluster's evolution (the red line in panel (c)).}
  \label{fractal_histories}
\end{figure*}

In the chosen simulation, $N_{\rm enrich} = 3$ G-dwarfs were enriched by the supernova. In Fig.~\ref{fractal_NN_dist} we show the distance to the nearest neighbour of each enriched G-dwarf as a function of time. One of the G-dwarfs is in a close ($\sim$10\,au) binary (the black line), and therefore is not a `singleton' and cannot be 
an analogue of our own Solar System. The remaining two enriched G-dwarfs have occasional encounters that are of order 500\,au. Encounters of this magnitude have been suggested as potential mechanisms to create the high eccentricities of some 
Edgeworth--Kuiper Belt Objects, such as Sedna \citep{Brasser06,Brasser12,Schwamb10}, and so the fact that the G-dwarfs in our simulations undergo sub--1000\,au encounters should not hinder planet formation and subsequent orbitial stability.

In Fig.~\ref{fractal_CC_dist} we show the distance from the cluster centre for each of the $N_{\rm enrich, sing} = 2$ singletons in the simulation. Due to dynamical mass segregation, the supernova progenitor has sunk to the cluster centre before the explosion, and both 
G-dwarfs are required to be on cluster-centric orbits to enable enrichment. This has two implications for our two enriched singletons. Firstly, they must pass through the cluster centre at least once -- during this time their discs could be subject to 
photoevaporation \citep[e.g.][]{Armitage00,Scally01,Adams04} from other massive stars, which are also likely to reside in the cluster centre. Inspection of Fig.~\ref{fractal_CC_dist} shows that one of our singletons, shown by the red line, passes though the inner 0.5\,pc of 
the cluster centre much more often than the other (shown by the green line). Secondly, at the time of the supernova (shown by the dotted vertical line in all panels), the singletons just happen to be passing through the cluster centre at that instant -- they are ``in the 
right place at the right time''.

Finally, we show the change in velocity magnitude for the enriched singletons in Fig.~\ref{fractal_Delta_V}. One of these singletons experiences velocity kicks in excess of 1\,km\,s$^{-1}$ (the red line), which we suggest could disrupt planet formation and/or evolution. 
We note that this enriched singleton also passes through the cluster centre more often, and therefore it is not surprising that it has a more hostile dynamical history than the other enriched singleton in this cluster.

\section{Discussion}
\label{discuss}

In Sections~\ref{results}~and~\ref{history} we have presented the results of $N$-body simulations of star cluster evolution in which we have investigated the numbers of Solar-type (G-dwarf) stars that could be enriched in short-lived isotopes by ejecta from the supernova of 
a 25\,M$_\odot$ star. In this scenario a supernova enriches the Sun's protoplanetary disc with the levels of $^{26}$Al and $^{60}$Fe found in meteorites from the epoch of planet formation \citep{Lee76,Adams11}. In order to experience the enrichment levels required without 
stripping too much of the disc away, the G-dwarf(s) must be within 0.1 -- 0.3\,pc of the supernova explosion \citep{Chevalier00,Ouellette07}. 

We have varied the initial conditions of the star cluster in an attempt to cover as much parameter space as possible for the initial conditions of star-forming regions that are likely to produce at least one 25\,M$_\odot$ star. In our first simulation, we adopt a smooth, virialised \citet{Plummer11} 
morphology; this model is unlikely to be representative of the initial conditions of star forming regions, which exhibit a high degree of substructure \citep[e.g.][]{Cartwright04,Sanchez09}. However, we expect the evolution of Plummer spheres to be less stochastic than fractal clusters 
\citep{Parker12b}, so these simulations provide a useful benchmark comparison to the fractal simulations.

In the remaining three suites of simulations we have created clusters with primordial substructure, and also varied the initial virial ratio. Observations of stars in star-forming regions have shown them to have subvirial (cool) velocities \citep[e.g.][]{Peretto06,Furesz08}, which 
in tandem with primordial substructure, facilitates a violent relaxation process resulting in a dense spherical cluster \citep{Allison10}. Such initial conditions have been successful in explaining the Orion Nebula Cluster, but do not lead to the formation of unbound associations. 
Indeed, massive unbound associations (e.g. Sco~Cen, Carina) have been suggested as the more likely birthplace of the Solar System, and such regions are observed to contain short-lived radioactive isotopes from nucelosynthesis \citep[e.g.][]{Diehl10,Voss12}. 
For this reason, we also ran simulations of substructured clusters with virialised (tepid) and supervirial (warm) velocities to investigate whether an unbound association could be a likely birthplace of the Sun \citep[see also][]{Pfalzner13}.

Somewhat surprisingly, the assumed initial virial ratio does not greatly affect the results. In the initially substructured models, 6/20 cool clusters and 5/20 tepid clusters contain enriched, unperturbed singleton G-dwarfs. The warm, 
substructured clusters also host a low number of potential Solar System analogues (6/20 clusters contain enriched unperturbed singletons). However, in this case the low densities achieved by these expanding associations are the cause of the low number, rather than hostile dynamical 
interactions. As an example, supervirial cluster number (xviii) in Table~\ref{warm_fractal_data} contains 0 enriched, unperturbed singletons. However, if we assume the supernova exploded at 4\,Myr instead of 6.64\,Myr, then the cluster contains 6 unperturbed singletons (from a total of 12 enriched G-dwarfs). 

Adding substructure to the clusters does appear to influence the results. In the Plummer models, we find that 15/20 clusters contain enriched G-dwarfs, but applying our constraints that the star must be a singleton and not suffer a perturbing velocity 
kick we find that 11/20 clusters contain several enriched, unperturbed singleton G-dwarfs. This is roughly a factor of two higher than the number of substructured clusters that contain enriched, unperturbed singletons, and is likely due to the fact that Plummer spheres are relaxed potentials, 
whereas the fractals take more than 1\,Myr to relax \citep{Allison10} and are therefore less quiescent environments. 

One interesting aspect of our results is that the numbers of enriched, unperturbed singletons in the clusters are rather uniform (albeit subject to low-number statistics) compared to the distribution of the number of enriched G-dwarfs. Taking two substructured, tepid clusters as an example, 
cluster number (i) in Table~\ref{tepid_fractal_data} contains 5 enriched G-dwarfs, 4 of which are unperturbed singletons, whereas cluster (ii) contains 15 enriched G-dwarfs, but only 3 unperturbed singletons. This highlights two processes within the numerical simulations; firstly,  the evolution 
of the clusters is highly stochastic (even the Plummer spheres contain a wide spread in $N_{\rm enrich}$) and secondly, large numbers of enriched stars equates to the supernova occuring in a particularly dense region. The higher chances of one of the 96 G-dwarfs experiencing enrichment  
in a dense cluster must be offset by the fact that the G-dwarfs will likely suffer more dynamical interactions and close encounters.

We caution that the number of enriched, unperturbed singletons from our simulations may be overestimated because we have assumed a field-like primordial binary fraction for G-dwarfs in the clusters. Current estimates suggest that the binary fraction in the field is $\sim$46\,percent \citep{Raghavan10} 
which may be significantly lower than the primordial binary fraction \citep[likely to be between 75 -- 100\,per cent, e.g.][]{Kaczmarek11,Parker11c,King12a}, due to dynamical processing. If we were to increase the binary fraction of the G-dwarfs, we would expect fewer enriched 
G-dwarfs to be `singletons' throughout the evolution of the cluster.

In addition to the supernova time, the other constraint for enrichment is that the supernova explodes between 0.1 and 0.3\,pc from the G-dwarf (a balance between stripping away too much of the protosolar nebula, and injecting enough radioactive isotopes into the disc). Panel (c) in  Figs.~\ref{plummer_cluster}~--~\ref{warm_frac_cluster} indicates that a relaxing of this upper bound would imply enrichment of more G-dwarfs; however, the boundary at 0.3\,pc already assumes highly efficient injection \citep{Ouellette07,Adams11} and it is unlikely that this is underestimated. 

Our simulations have shown that -- whilst it is possible to have multiple enriched, unperturbed singletons in a cluster -- 50 to 75\,per cent of the clusters we model do not contain any. Furthermore, the late-stage injection of $^{26}$Al and $^{60}$Fe, which occurs at 6.6\,Myr for a supernova with 
progenitor mass 25\,M$_\odot$, has been cited by \citet{Gounelle08,Gounelle12} as being too late in the disc evolution to be homogeneously included in the meteorites. This could in principle be alleviated if the supernova progenitor formed first, and the lower-mass stars formed several Myr later, although 
this would require firm evidence of age spreads in star-forming regions. Evidence for and against such  age spreads is currently the subject of much debate \citep[e.g.][]{Reggiani11b,Jeffries11}. Recently, however, \citet{Bell13} suggested that the ages of star-forming regions and open clusters may be 
underestimated by a factor of two; if this is the case, then protoplanetary discs are longer-lived and the arguments against late-stage injection become weaker.

We are then left with the straightforward question: is late-stage enrichment too uncommon to be a feasible delivery mechanism for the $^{26}$Al and $^{60}$Fe levels in the early Solar System?  Models of Solar System formation triggered by a supernova naturally account for the 
$^{26}$Al \citep{Cameron77,Boss95,Gritschneder12,Vasileiadis13}, but require that $^{60}$Fe abundances in the early Solar System be similar to the background ISM levels \citep[as recently suggested by][]{Moynier11,Tang12}. Indeed, \citet{Gounelle09,Gounelle12} recently 
proposed a three-stage star formation scenario where the supernovae of several stars produces the $^{60}$Fe, then triggers the formation of a massive star which produces $^{26}$Al during its Wolf--Rayet phase. The Sun is then born in a third generation of star formation with the correct enrichment levels 
\citep[see also][]{Gaidos09}. 
Such a scenario would require age spreads to be observed in star-forming regions (like the disc enrichment model), but of the order $\sim$15\,Myr, and is perhaps more unlikely than the age spread of $\sim$3\,Myr that would alleviate drawbacks in the disc enrichment model.

If we assume that all the clusters in our simulations eventually evaporate into the Galactic field \citep[and that dynamical interactions after 10\,Myr are so infrequent as to not affect our results,][]{Dukes12,Pfalzner13}, then the fraction of enriched, unperturbed singletons can be derived by dividing 
the sum of $N_{\rm enrich, sing, unp}$ from all clusters by the sum of all G-dwarfs (i.e.\,\,96 $\times$ 20 simulations, a total of 1920 G-dwarfs). The sum of $N_{\rm enrich, sing, unp}$ ranges from 19 (the smooth, tepid clusters) to 6 (the substructured, warm clusters). The fraction is therefore of 
order 1\,per cent (smooth, tepid), or 0.3\,per cent (substructured, warm) and of order 0.5\,per cent for the substructured, cool/tepid clusters. If we relax the constraint that the enriched singletons must not suffer velocity perturbations, the enriched singleton fraction is between 2.2 (smooth, tepid) and 0.4\,per cent (substructured, warm).

Finally, we note that the dynamical evolution of star clusters is highly stochastic. Some clusters eject the supernova progenitor before enrichment, and even when enrichment occurs the numbers of enriched singleton G-dwarfs varies between 0 and 7, simply through subtle differences in the evolution 
of our (initially) statistically identical model clusters. Such stochasticity is impossible to characterise observationally without prior information on the star formation process \citep{Parker12b}.

\section{Conclusions}
\label{conclude}

Evidence of short-lived isotopes in meteorites suggests that the Sun was in close proximity to a 25\,M$_\odot$ star which went supernova at the epoch of planet formation in the Solar Sytem. If these isotopes are delivered to the Sun's protoplanetary disc, the supernova must 
have occurred at a distance between 0.1 -- 0.3\,pc of the Sun. We have conducted $N$-body simulations of the dynamical evolution of star clusters with $N = 2100$ members, which would be expected to form at least one 25\,M$_\odot$ star under the assumption of a normal IMF. 
We have determined the number of G-dwarfs that experience the necessary levels of enrichment, and then determined their dynamical histories to ascertain whether they could potentially be Solar System analogues. Our conclusions are the following:

(i) Typically, between 50\,per cent and 75\,per cent of clusters contain supernova-enriched G-dwarfs at a distance of between 0.1 and 0.3\,pc from the supernova. The number of enriched G-dwarfs is in the range of several to over 10 (from a total of 96). 

(ii) If we consider only `singletons' -- G-dwarfs that are never in close binaries, or suffer disruptive encounters, then only $\sim$25\,per cent of clusters contain G-dwarfs that are enriched, unperturbed singletons. Usually these clusters contain only one or two such objects. 

(iii) The assumed initial conditions for star formation have little impact on the results; there is little difference in the numbers of enriched, unperturbed singletons between substructured clusters which are subvirial (cool collapse), virial (tepid and static) or supervirial (warm and expanding). The 
only caveat is that if the supernova were to explode earlier, the expanding supervirial clusters would have a higher occurrence of enriched, uperturbed singletons.

(iv) Summing together all the G-dwarfs from each suite of simulations, the global fraction of G-dwarfs that are enriched, unperturbed singletons is of order 0.5 -- 1\,per cent.

(v) The cluster evolution and numbers of enriched stars is highly stochastic; statistically identical clusters can enrich over 10 G-dwarfs, or only several, or none at all -- differences are due to the inherently chaotic nature of star cluster evolution.\\

At first sight the $N$-body models suggest that supernova enrichment of unperturbed singleton G-dwarfs like our Sun is a rare occurrence. However, it does occur in a significant fraction of clusters, and sometimes to more than one G-dwarf in the same cluster. Future investigation of the assumed cluster parameters (IMF, morphology, density, virial 
state), and the time of the supernova explosion, would be beneficial to investigate whether the probability of disc enrichment can be raised by changing one or more of these parameters.  

\section*{Acknowledgments}

We thank the anonymous referee for their comments and suggestions, which improved the original manuscript. RPC was supported by the Swedish Research Council (grant 2012-2254). MBD was supported by the Swedish Research Council (grants 2008-4089 and 2011-3991). The simulations in this work were performed on the \texttt{BRUTUS} computing cluster at ETH Z{\"u}rich.

\bibliographystyle{mn2e}
\bibliography{general_ref}

\label{lastpage}

\end{document}